\documentclass[floatfix,pre,twocolumn,english]{revtex4}
\usepackage{graphicx}
\usepackage{amsmath}
\usepackage{amssymb}
\usepackage{color}
\makeatletter

\usepackage{psfrag}

\usepackage{babel}
\newcommand\Pf{\operatorname{Pf}}

\makeatother

\begin{document}
\newlength{\singfigwidth}
\setlength{\singfigwidth}{0.9\columnwidth}
\newlength{\doublefigwidthone}
\setlength{\doublefigwidthone}{6.5in}
\newlength{\doublefigwidth}
\setlength{\doublefigwidth}{6.5in}

\title{Exact Algorithm for Sampling the 2D Ising Spin Glass}
\author{Creighton K. Thomas}
\author{A. Alan Middleton}
\affiliation{Department of Physics, Syracuse University, Syracuse, NY 13244, USA}

\begin{abstract}
  A sampling algorithm is presented that generates spin glass
  configurations of the 2D Edwards-Anderson Ising spin glass at finite
  temperature, with probabilities proportional to their Boltzmann
  weights. Such an algorithm overcomes the slow dynamics of direct
  simulation and can be used to study long-range correlation functions
  and coarse-grained dynamics.  The algorithm uses a correspondence
  between spin configurations on a regular lattice and dimer (edge)
  coverings of a related graph: Wilson's algorithm [D.~B.~Wilson,
    Proc.\ 8th Symp.\ Discrete Algorithms 258, (1997)] for sampling
  dimer coverings on a planar lattice is adapted to generate samplings
  for the dimer problem corresponding to both planar and toroidal spin
  glass samples.  This algorithm is recursive: it computes
  probabilities for spins along a ``separator'' that divides the
  sample in half. Given the spins on the separator, sample
  configurations for the two separated halves are generated by further
  division and assignment.  The algorithm is simplified by using
  Pfaffian elimination, rather than Gaussian elimination, for sampling
  dimer configurations.  For $n$ spins and given floating point
  precision, the algorithm has an asymptotic run-time of $O(n^{3/2})$;
  it is found that the required precision scales as inverse
  temperature and grows only slowly with system size. Sample
  applications and benchmarking results are presented for samples of
  size up to $n=128^2$, with fixed and periodic boundary conditions.
\end{abstract}
\maketitle

\section{Introduction}
Materials with quenched disorder, such as spin glasses, can have
extremely long relaxation times, so that laboratory samples exhibit
non-equilibrium behavior over many decades in time scale
\cite{BinderYoung,YoungBook,MemoryExpt}.  Spin glass materials exhibit
``aging'', a slow evolution in the magnetic response, for example, and
non-equilibrium phenomena such as ``rejuvenation'', where changes in
the temperature can undo the effects of aging.  As these phenomena
take place over time scales much longer than the microscopic time
scale for individual spins, these effects must be due to the
collective behavior of many spins.  As analytical work is very
difficult in disordered materials \cite{FisherHuse,ParisiRSB},
numerical simulations have been important in building a picture of the
low-temperature phase of models of disordered spin systems (e.g.,
\cite{KatzgraberYoung,Rieger,Cugliandolo}).

Numerical work using direct local Monte Carlo simulation of the
dynamics and equilibration \cite{JANUS} indicate that models
such as the Edwards-Anderson model \cite{EA} possess the long
relaxation times that are at least necessary to start to explain these
behaviors.  Given the direct correspondence between simulation time
and ``experimental'' time, though, the same long relaxation times that
one is seeking to understand make such simulations very difficult,
even though very long simulation times are used \cite{JANUS}.

Various alternate approaches and approximations have been developed to
address the difficulties of direct simulation. These approaches can be
used to determine both the equilibrium state and how this state is
approached. When the primary concern is the understanding of the
equilibrium state, many studies have sought to find the ground state
of given samples, as many of the properties of the low-temperature
phase are believed to be given by the properties of the ground state
(such as the sample-to-sample fluctuations in the ground state energy
or the length-dependent domain wall free energy)
\cite{HoudayerMartin,ManyHartmann,LiersEtal,PalassiniYoung,BoettcherEO}.
This direction of research is based on developing faster exact methods
and accurate heuristic methods for finding the spin configuration that
minimizes a Hamiltonian with fixed random couplings. The search for a
ground state configuration is closely connected with combinatorial
optimization methods developed in computer science, though
finite-dimensional spin glasses additionally lend themselves to
real-space techniques inspired by the renormalization group
\cite{HoudayerMartin}. Equilibrium quantities at finite temperature,
such as the partition function and density of states, can be computed
for the 2D Ising spin glass. The approach to the ground state and
non-equilibrium properties can then be studied by direct simulation or
possibly heuristically by real-space blocking of the degrees of freedom
\cite{PatchworkDynamics}.

We present here an algorithm that extends these approaches to allow for exactly
sampling the configurations of the disordered Ising model on 2D lattices
without the use of Markov Chain Monte Carlo (MCMC). For $n$ spins, this
algorithm takes $O(n^{3/2})$ steps and in practice has a running time that
grows only somewhat faster, i.e., somewhat more rapidly than $L^3$, at fixed
temperature. As lower temperatures $T$ require more precise arithmetic, the
running time grows roughly as $T^{-1}$. The algorithm is based on Wilson's
algorithm for sampling planar dimer models \cite{Wilson}. We use a mapping of
the Ising spin glass model to the dimer problem for the decoration of the graph
dual to the spin lattice \cite{Barahona,KastCities}. We take advantage of the
regular structure of the square lattice to simplify the algorithm and also
modify the matrix algebra of Wilson's algorithm so that the calculation is both
simpler and more numerically stable.

This algorithm for sampling provides an opportunity to study many outstanding
questions for 2D spin glasses in much more detail than possible with MCMC
computations. For example, the dependence of replica overlaps on temperature
and sample size can be directly computed. Correlation functions are easily
found: these can be used to study the decay of correlations at finite
temperature in both Gaussian and $\pm J$ models, which differ in some aspects
at $T=0$. The power law decay of spin-spin correlations are presumed to behave
as $r^{-\eta}$ up to the correlation length: how $\eta$ depends on model and is
related to thermodynamic quantities such as the heat capacity is still not
completely understood \cite{correlations}.

\subsection{Model}

The Edwards-Anderson (EA) spin glass model is a prototypical model for
disordered materials.  The EA spin glass model has the Hamiltonian
\begin{eqnarray}
  {\mathcal H}_\mathcal{J}(\mathcal{S}) & = & -\sum_{\langle ij \rangle} J_{ij} s_i s_j,
\end{eqnarray}
where the $\mathcal{J}=\{J_{ij}\}$ are sample-dependent couplings.
For example, the $J_{ij}$ can be chosen independently and randomly
from a Gaussian distribution or from a bimodal distribution
$J_{ij}=\pm 1$ (the $\pm J$ model), with mean zero and variance 1 in
either case. These couplings connect two neighboring spins, located at
points $i$ and $j$ in the sample. The spins $s_i$ are Ising spins,
i.e., each $s_i=\pm 1$.  We will only be able to exactly sample in the
2D case. We will study the square lattice of spins in both the case of
periodic boundary conditions, where the bottom row of spins is
connected to the top and the left column to the right column, and the
case of fixed boundary conditions, where the spins on the boundary of
the square sample are fixed.  A spin configuration
$\{s_i\}=S\in\mathcal{S}$ is an assignment of spin values $s_i$ to
each of $n$ sites $i$; there are $2^n$ possible spin configurations in
the state space $\mathcal{S}$. A ground state spin configuration
$S_{\mathrm GS}$ that minimizes the Hamiltonian can be found in
polynomial time using a minimum-weight perfect matching algorithm, if
the edges $\langle ij \rangle$ which connect nearest neighbor sites
and the sites $\{i\}$ form a planar graph \cite{Barahona}. At positive
temperature $T=\beta^{-1}$, the partition function for a given
realization of disorder $\mathcal{J}$ is $Z_{\mathcal{J}}=\sum_{S'}
\exp[-\beta{\mathcal H}_\mathcal{J}(S')]$ and the probability of
observing a spin state $S$ in a sample defined by $\mathcal{J}$ is
$P_{\mathcal{J}}(S)=Z_{\mathcal{J}}^{-1} \exp[-\beta{\mathcal
    H}_{\mathcal{J}}(S)]$ in equilibrium.

\subsection{Exact computation of the partition function}
It has long been known that the partition function of the 2D
ferromagnetic ($J_{ij}\equiv 1$) Ising model with no external magnetic
field can be found exactly by computing the determinant of a matrix
derived from the spin lattice.  One type of construction of this
determinant uses a sum over sets of closed loops on the spin lattice:
these loops represent the terms in a high-temperature expansion of the
partition function.  The first published construction of these type of
loops is that of Kac and Ward \cite{KacWard}, who directly count the
polygonal loops.  A technique for constructing the relevant matrix for
the determinant technique is to map the Ising model onto a dimer
covering problem on a decorated lattice $G$
\cite{KastCities,FishCities}, where the spins in the original lattice
are replaced by a subgraph, a Kasteleyn or Fisher city (a dimer
covering is a set of edges in the graph such that every node belongs to
exactly one selected edge).  The Kasteleyn matrix $K$ of the graph $G$ for the
dimer problem describes the connections between neighboring nodes.
This square matrix, which is indexed by a numbering of the nodes of
$G$, has non-zero entries at locations that are indexed by the two ends of
a connection between the nodes.  Counting the partition function for
dimer coverings is equivalent to computing the Pfaffian of the
Kasteleyn matrix, where the Pfaffian in this case is a square root of
the determinant.  These Pfaffian techniques have been used for the
exact solution of the pure Ising model in the thermodynamic limit
\cite{KacWard,KastCities,FishCities} and, e.g., for computing the
density of states in finite samples. Beale \cite{Beale} rewrote the
Pfaffian in a form that allows for faster direct computation of the
partition function in a pure ferromagnetic model. As the derivation of
the correspondence between the partition function of the Ising model
and the determinant or Pfaffian methods for finite samples does not
rely on a homogeneous coupling constant $J_{ij}$, these methods can
also be applied to spin glass samples in two dimensions.  This
correspondence has thus been used to compute directly the partition
function (and density of states) for disordered samples
\cite{SaulKardar,GLV}.  Pfaffian techniques can also be used to
compute degeneracies and correlation functions in the $\pm J$-model
(where couplings are all of the same magnitude, but randomly
ferromagnetic or antiferromagnetic between neighboring spins)
\cite{BlackmanPoulter} and has been used to study the heat capacity of
this same model at low temperatures (e.g., see \cite{LukicEtAl}).

\subsection{Review of configuration sampling}

Being able to compute the partition function (and often the density of
states as a by-product) is useful in computing such quantities as
domain wall free energies, sample-to-sample fluctuations in the free
energy, specific heat, and other global quantities.  By computing the
partition function for fixed relative spin configurations, one can
also calculate correlation functions \cite{BlackmanPoulter}. But for
many purposes, such as faster computation of correlation functions,
the organization of states in a spin glass, or for use in a heuristic
for studying the dynamics of disordered materials
\cite{PatchworkDynamics}, it is useful to be able to generate sample
configurations, given a realization of the disorder.  For sampling the
equilibrium behavior of the system, it is sufficient to generate such
samples with their proper Boltzmann probability
$P_{\mathcal{J}}(\mathcal{S})$.  For nonequilibrium dynamics, such
sampling can be used in patchwork dynamics, which is closely related
to the renormalization approaches to nonlocal dynamics used in
multigrid Monte Carlo methods and hierarchical genetic methods
\cite{HoudayerMartin,Janke}.

Heuristic sampling, where there is no proof of exactness, is typically
done using the Markov chain Monte Carlo (MCMC) method. In MCMC
methods, local probabilistic dynamics that obey detailed balance are
used to update the spins.  At long times, the probability of observing
a configuration should be the equilibrium probability. The
equilibration times using this method can be prohibitively long,
though, especially in glassy systems such as the 2D spin glass
\cite{2DISGMonteCarlo}.  Some faster Monte Carlo methods
have been developed for the
2D spin glass at low temperature \cite{FasterMCAlgs},
but with any such method there is also a question of how to test
whether equilibrium is achieved with sufficient accuracy. It is of use
to have criteria to confirm converges of the Markov chain to the
equilibrium distribution.  Propp and Wilson \cite{ProppWilson}
proposed a technique for generating \emph{exact} samples with MCMC by
``coupling from the past'' (CFTP). In this framework, it is possible
to verify that the system has converged from all possible initial
conditions to a single state, at which point it is exactly in
equilibrium.  This approach often makes use of a natural partial
ordering of configurations that is used to guarantee convergence.  For
disordered models, there is often no such obvious partial ordering of
the states that ensures convergence of CFTP.  Chanal and Krauth
\cite{ChanalKrauth} have nevertheless succeeded in applying CFTP to the Ising
spin glass using a coarse-grained organization of the states: at
first, all states are possible; as the Markov chain is developed and
the number of states is reduced by coupling, the constraint on allowed
states is further coarse-grained, until a single whole sample state is
left.  But the coupling time (time for convergence to a single sample)
is still of the order of the equilibration time, which of course can
be very long at low temperatures.

Sampling with the exact Boltzmann weights has been implemented and applied to
the Migdal-Kadanoff (MK) lattice, which is not a finite-dimensional lattice,
but is used to approximately represent finite-dimensional lattices. As the MK
lattice has a hierarchical structure, the spin configurations can be summed
over successive scales, starting from the smallest, to compute the partition
function and the relative partition functions can be used to sample the spins.
This was done in Refs.\ \onlinecite{SasakiMartin} and
\onlinecite{JorgKatzgraber} to study chaos and spin overlap on hierarchical
lattices.

Exact sampling of configurations can always be carried out in time
polynomial in the size of the sample, if the partition function may be
calculated efficiently.  One direct, but somewhat slow method, is to
assign a single spin at random and then compute the partition function
conditioned on assignment of individual neighboring spins; this
requires $n=L^d$ computations of the partition function for $O(n)$
spins.  Such a technique is mentioned as a possibility, for example,
in Ref.\ \onlinecite{KrauthBook}. As the partition function can be
computed in $O(n^{3/2})$ steps, this would require $O(n^{5/2})$
arithmetical steps. There are other methods for carrying out exact
sampling, however.

Exact sampling of ferromagnetic Ising systems (in any dimension) may
be performed in polynomial time \cite{RandallWilson}.  This technique
works in the Fortuin-Kasteleyn cluster representation and successively
removes bonds and spins through a reduction technique.  A related
problem, sampling configurations of dimer coverings on a planar
bipartite lattice, has an elegant sampling technique
\cite{ProppAD,ZengEtal}, which exactly maps the statistical mechanics
on an $L\times L$ lattice to an $(L-1) \times (L-1)$ lattice with
modified weights on the edges. Other techniques for calculating the
exact partition function of the 2D Ising Spin Glass, such as the
Y-$\Delta$ technique of Loh and Carlson \cite{LohCarlson}, are quite
similar in spirit to the dimer covering algorithm.  This technique
also involves an efficient recursive reduction of any planar graph to
a smaller graph, but when frustration is present the intermediate
reduced bond strengths can become complex, which complicates possible
sampling techniques.

\subsection{Overview of algorithm}

We now outline the crucial points for our application.  In two
dimensions, there is a one-to-one correspondence between spin
configurations of the Ising model with arbitrary couplings and dimer
configurations on a decorated version of the dual lattice. The
individual spin and dimer configurations have the same energy, so the
corresponding configurations have the same Boltzmann weights
$Z^{-1}\exp(-\beta E)$, where $Z$ and $E$ are the partition function
and configuration energy for either the dimer or spin problem.  We can
therefore generate sample spin configurations by sampling among dimer
configurations and mapping them to the spin representation.  Note that
the traditional method for calculating the partition function is a
mapping between the primal lattice and a dimer model: a dimer
configuration, which defines loops in a high temperature expansion of
the partition function, does not directly map onto a unique spin
configuration. Using the dual lattice, however, allows for such a map.

Wilson's algorithm may be used to sample dimer configurations
efficiently for any planar lattice, so efficient sampling of the Ising
model can be carried out on general planar samples. One requirement
for Wilson's algorithm is an efficient method to recursively subdivide
the lattice; this task is straightforward on a regular lattice: we
subdivide or separate the sample by choosing two adjacent rows or
columns of spins. The spins on these two lines are the separator sites
for the spin lattice.  These separator spins are then assigned by a
sequence of weighted choices. The weights for the choice of these
spins are found, in essence, by computing the needed correlators
between each pair of spins situated on these two lines.  Once the
spins on the separator have been chosen and fixed, this division and
sampling is repeated on finer and finer spatial scales, using the
solved spins as fixed boundary conditions for the subsamples.  Besides
allowing for recursive assignment of spins on the separators, this
nested dissection is used to efficiently organize the needed sparse
matrix computations.

We have also simplified the algorithm significantly by using Pfaffian
elimination, rather than Gaussian elimination.  Pfaffian elimination
was used by Galluccio, Loebl, and Vondrak \cite{GLV} in computing the
partition function, but it can also be used to advantage in sampling.
We use a sparse matrix representation that greatly reduces the
amount of space and time needed: due to the regular nature of the
lattice, all of the primitive operations can be explicitly precomputed
and then applied to many distinct samples of the same size. We find
that the number of relevant matrix elements (out of the full $O(n^2)$
potential elements) that are ``visited'' during the computation scales
approximately $\sim n$ and that the number of operations obeys the
expected growth $\sim n^{3/2}$.

Though the form of the algorithm that we use is based upon and
parallels Wilson's algorithm, we present the method in detail here. We
do this in order to review the method itself, emphasize 
the relationship between
matchings and the Ising model, present our form of the matrix
algebra that we use for sampling dimer matchings, and describe
sampling for non-planar graphs, such as used for periodic boundary
conditions.

\subsection{Implementation results}

\begin{figure*}
  \includegraphics[%
    width=\doublefigwidthone, keepaspectratio]{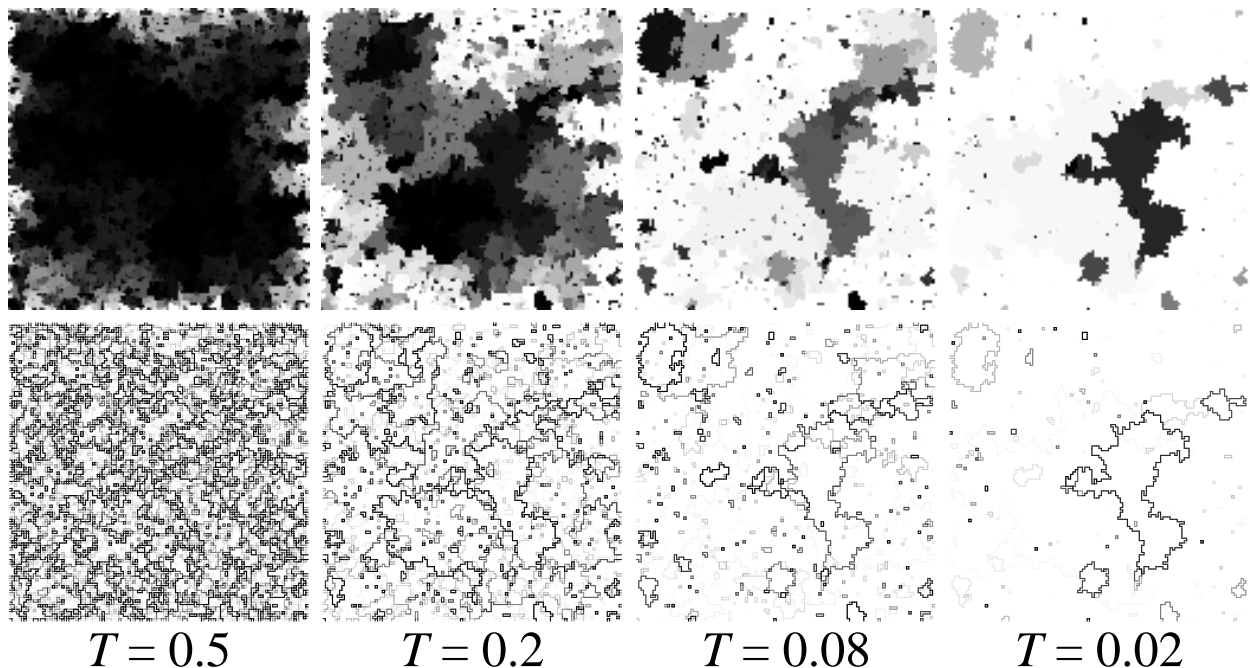}
  \caption{\label{fig:firstFish} Results of applying the sampling
    algorithm to an individual 2D Ising spin glass sample, for
    temperatures $T=0.5,0.2,0.08,0.02$, for a single Gaussian spin
    glass sample with fixed boundaries.  The images show the variability of the spin
    assignments (top) and of the domain walls (bottom) over a range of
    temperatures, in a sample with $n=126^2$ variable spins surrounded
    by a layer of fixed spins.  At least 240 samples were generated at
    each temperature. The gray scale values indicate the probability
    of a given spin being fixed (upper row) or of neighboring spins
    being fixed relative to each other (lower row).  For spin
    assignments, the darkest colors indicate that the spin is equally
    likely to be up or down, while light colors indicate that the spin
    occurs with a single alignment in nearly all sampled
    configurations.  These alignments result from correlations with
    the fixed boundary spins.  For the domain walls displayed in the
    lower row of images, the lines indicate the probability of
    relative domain walls between two configurations: the darkest
    lines indicate the bond dual to that domain wall has a 50\% chance
    of opposite or equal relative orientations; where there is no line
    separating two spins (or only a very light one), the two spins
    have a very high probability of a single relative orientation,
    either aligned or opposite.  Specifically, the bond satisfaction
    variance $\mu_{i,j}(1-\mu_{i,j})$ is plotted along
    each dual edge, where $\mu_{i,j}$ is the frequency of the
    $J_{ij}s_is_j$ being positive. Note that as $T$ decreases,
    the frequency of specific droplet excitations, outlined by domain
    walls, can either increase or decrease, reflecting the sensitivity
    of the configurations to temperature. This can be seen, for
    example, in two of the regions that are active at $T=0.02$, the
    approximately $20\times20$ region in the far upper left and the
    approximately $30\times60$ region at the center right: the spins
    in the former become more fixed as temperature decreases while the
    spins in the latter region become more variable (darker) when the
    temperature is decreased from $T=0.08$ to $T=0.02$.  }
\end{figure*}

As one of the primary motivations for the development of our algorithm
is its potential use in patchwork dynamics \cite{PatchworkDynamics},
we test our algorithm by timing it in this context, random patches of
a sample with Gaussian bonds, where the variance of the couplings
$J_{ij}$ is unity and the mean coupling $\overline{J_{ij}}=0$. Our
code was developed with the possibility of using different data types
as the matrix elements in the calculation. Specifically, we test the
algorithm using double precision numbers, floating point numbers of
arbitrary precision, and with exact rational Boltzmann weights.  As
the weights in the computation can vary over a large range and a Pfaffian
elimination technique is used to cancel out matrix elements, similar
to Gaussian elimination, the algorithm can produce unstable results
using the floating point types, if proper care is not taken.  The
likelihood of an instability increases with increasing system size and
with lower temperature.  In trying to balance the stability and
accuracy of the sampling against the running time, we determine the
arithmetical precision needed to reliably sample a
configuration.
Sample
results for configurations are displayed in Fig.\ \ref{fig:firstFish}.
Details of the precision requirements and example running times are
given in Sec.\ \ref{subsec:typesprecision}.

\begin{figure}
  \includegraphics[%
    width=\singfigwidth, keepaspectratio]{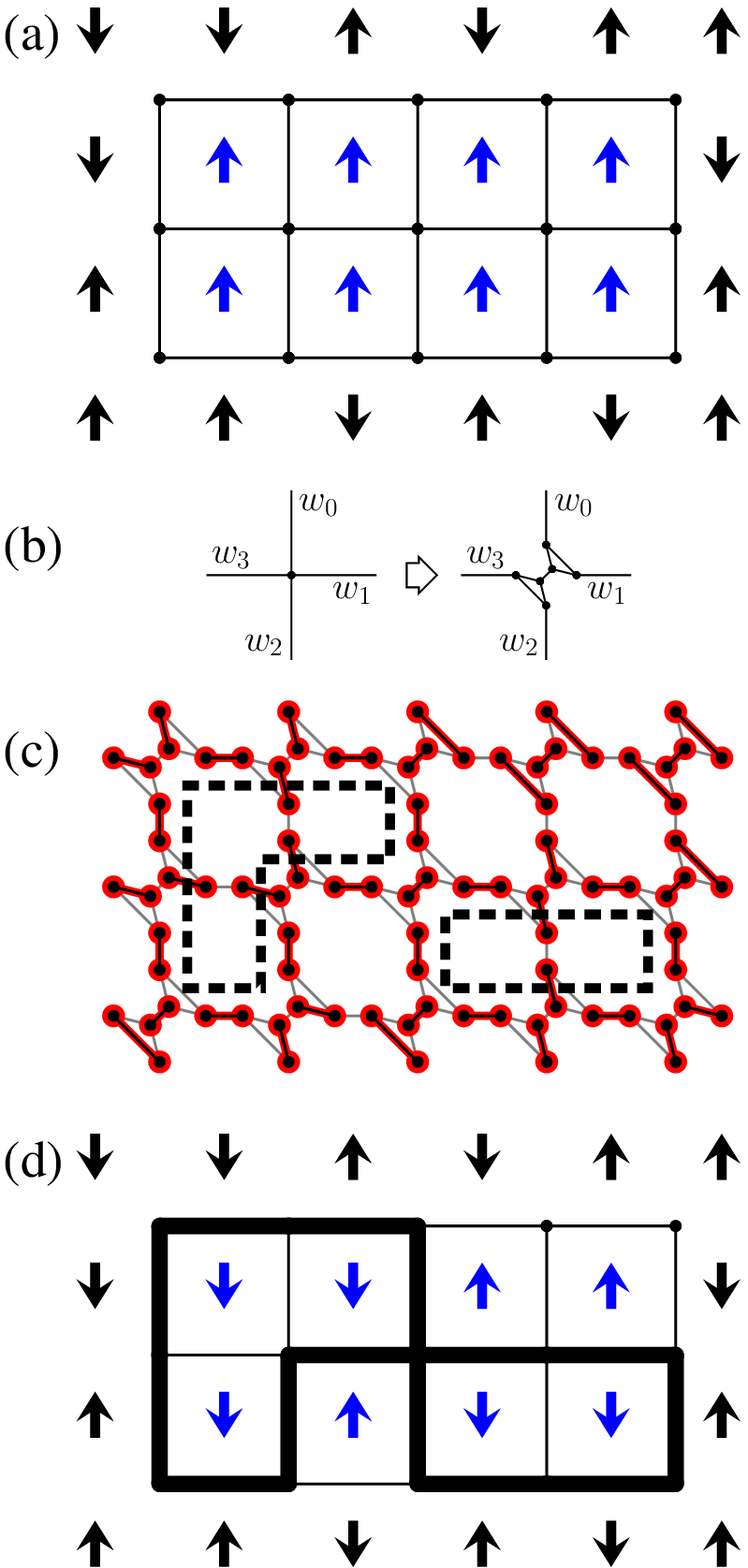}
  \caption{\label{fig:dualLoopsSpins} {[}color online{]} A depiction
    of the correspondence between domain wall loops for an Ising spin
    system and dimer matchings on the decorated dual lattice $G$.  (a)
    A spin system with fixed boundary conditions; an up arrow at location $i$ indicates $s_i=+1$
    and a down arrow indicates $s_i=-1$.  The dual lattice $G_0$ is indicated
    by the lines connecting the dual nodes.  (b) A Fisher city
    replacement.  Each dual lattice node is expanded to a Fisher city,
    a set of six nodes composed of two linked triangles, to generate
    the decorated lattice.  For work on the square lattice, the bond
    strengths are set to be $w(e_{ij})=0$ inside the city, and the
    bond strengths between the cities, indicated here by the notation
    $w_d$, $d=0,1,2,3$, are set according to Eq.\ (\ref{eqnWeights}).
    (c) An example dimer covering (i.e., perfect matching) $M$ on the
    decorated graph $G$. The thicker (also red) bonds with circular
    ends indicate
    edges in $M$.  The domain walls, composed of dimers that connect
    distinct cities, are indicated by dashed lines.  (d) When the
    cities are contracted out from $G$, the loops on $G_0$ remain.
    Given this choice of dimer covering $M$, the spins that are inside
    the domain walls are flipped to create the new sampled
    configuration.}
\end{figure}

\section{Mapping the Ising model to a dimer model}\label{Sec:Mapping}

In order to sample Ising spin configurations via the sampling of dimer
configurations, one requires a one-to-one correspondence between the
Ising spin configurations $\mathcal{S}$ on a given lattice and the
dimer covering configurations $\mathcal{M}$ on a related graph $G$.
Such mappings have been constructed for application to the more
straightforward problem of computing the partition function. These
mappings link the problem of computing $Z_{\mathcal{J}}$ to a weighted
enumeration of all perfect matchings $\mathcal{M}$ on $G$.  A single
perfect matching on a graph $G=(V,E)$, where $V$ are vertices (nodes)
and $E$ are edges connecting pairs of vertices, is a choice of a
subset of edges $M\subset E$, the matching or dimer covering, such
that every vertex belongs to exactly one edge in $M$ (see
Fig.\ \ref{fig:dualLoopsSpins}).  The generally established procedure
for constructing a mapping between spin configurations and perfect
matchings is to identify closed loops on some relevant graph, $G_0$,
where $G_0$ is either the primary grid (the spin lattice) or the dual
lattice (the lattice of plaquettes).  The partition function,
originally a sum over spin configurations, can be represented as a
weighted sum over choices of loops in $G_0$.  This summation over
loops can be carried out by summing over matchings on a graph $G$,
constructed by replacing the nodes of $G_0$ with either Kasteleyn or
Fisher ``cities'' \cite{KastCities,FishCities}, subgraphs constructed
of a few nodes and edges. Perfect matchings on this decorated lattice
$G$ then have the property that an even number of the covered edges
are incident upon any given city. The edges of a matching $M$ that
connect cities are therefore even at each city; contracting the cities
back to single points then gives the city-connecting dimers that compose
the loops in $G_0$ (see Fig.~\ref{fig:dualLoopsSpins}).

One mapping between spin configurations and sets of loops is based on
a high temperature expansion of the partition function of the Ising
model, where $G_0$ is the spin lattice and the loops, composed of
bonds connecting nearest-neighbor spins, represent individual terms in
the expansion of $Z_{\mathcal J}$ in powers of $\exp(-\beta J_{ij})$.  The
direct replacement of each Ising spin with a ``city'' gives
representation of loops by a dimer matching
\cite{KastCities,FishCities,GLV}.  The weight of dimer configurations
can then be summed using Pfaffian methods \cite{KastCities} giving,
for example, the Kasteleyn solution of the Ising model.  However,
there is no direct correspondence between individual sets of loops and
spin configurations.

Alternately, a mapping to $G$ can be defined by taking $G_0$ to be the
dual lattice \cite{Barahona,GSKastCities}.  This mapping, in contrast
with the approach of decorating the original lattice, allows for
direct sampling of Ising spin configurations.  The loops on the dual
graph represent a loop expansion in terms of domain walls.  The
expansion in domain walls, if expressed relative to the ground state,
would be a low-temperature expansion. More generally, the summation is
over relative domain walls between a reference configuration and any
other configuration. A direct correspondence between spin
configurations and dimer configurations therefore exists as domain
walls uniquely define a spin configuration, given a reference
configuration, up to the possibility of a global spin-flip symmetry.

Let $R=\{r_i\}$ be a reference configuration of Ising spins $r_i=\pm
1$. We emphasize that this choice is completely arbitrary: it need not
be a ground state. For convenience $R$ can be a configuration with all
spins up or a previously sampled configuration.  For a given sampling
$S$ of the spin configuration, $S=\{s_i\}$, the loops of dual edges
that separate spins $i$ and $j$ with $r_i s_i \neq r_j s_j$ define the
relative domain walls between $R$ and $S$.  (For the ferromagnetic
Ising model, one usually takes $r_i\equiv 1$, so that the domain walls
separate regions where $s_i=1$ from regions where $s_i=-1$.)

In this reference configuration, for each pair $i,j$, define $R_{ij} =
r_i r_j$ as the reference satisfaction of bond $i,j$.  Then, for this
fixed $R_{ij}$, we can simply rewrite the Hamiltonian as
\begin{eqnarray}
  \mathcal{H_J}(S) & = & - \sum_{\langle ij \rangle} J_{ij} \left( s_i s_j
  - R_{ij} + R_{ij}\right)\nonumber\\
  & = & \mathcal{H}_R + \mathcal{H}_G\,,
  \label{Eq:spin_to_dimer}
\end{eqnarray}
with $\mathcal{H}_R = - \sum_{\langle ij \rangle} J_{ij} R_{ij}$, the
energy of the reference configuration and $\mathcal{H}_G = -
\sum_{\langle ij \rangle} J_{ij} \left( s_i s_j - R_{ij} \right)$,
which will be rewritten as the Hamiltonian of the corresponding dimer
model is the energy of the domain walls between the configurations $R$
and $S$.  Note that $\mathcal{H}_R$ is the same for all spin
configurations, but must be tracked if comparing the effects of
changing boundary conditions or comparing with ground state energies,
for example.

Let the decorated graph $G=(V,E)$ have the vertex set $V$, which has
size $|V|=2N$, $N$ being the number of dimers in a perfect matching of
the vertices, and the edge set $E=\{e_{qr}\}$ where each edge
connects two nodes, $e_{qr}=(q,r)$, for some $q,r\in V$. Then, given
a set of relative domain wall loops, the dimer configuration is
uniquely defined by selecting dimers that connect cities and cross
bonds $J_{ij}$ where $s_i s_j \neq R_{ij}$, i.e., that overlie the
domain walls in $G_0$, and the subsequent unique choice of matching
for dimers internal to the cities.  Choosing an energy function $w(e)$
for edges in $E$ with $w(e_{qr})=0$ for bonds in the cities and
\begin{equation}
  w(e)=2 J_{ij}R_{ij} \label{eqnWeights}
\end{equation}
for dual edges $e$ crossing bonds between spins $i$ and $j$ gives
\begin{eqnarray}
  \mathcal{H}_G(M) & = & \sum_{e \in M} w(e) \label{eqnEnergyM}
\end{eqnarray}
as a consistent energy function for matching configurations in $M$.
The Ising model and matching model can therefore be made equivalent,
up to a global energy shift $\mathcal{H}_R$.

\begin{figure}
  \includegraphics[%
    width=\singfigwidth,
    keepaspectratio]{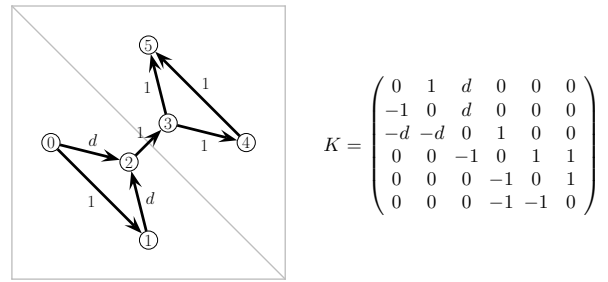}
  \caption{\label{fig:fish_city_mtx} The node indexing and edge
    orientations within a Fisher city (left) and the corresponding
    elements of the $6\times 6$ submatrix of the Kasteleyn matrix $K$
    (right).  The numbering of nodes is shown for the first city
    listed in the dual lattice; subsequent cities have multiples of 6
    added to their indices.  In the case of the square spin lattice
    (indicated by the outer square bonds on the original spin
    lattice), all non-zero $K$ elements, including $d$, are set to
    unit magnitude.  The labeling of the $0\rightarrow 2$ and $1
    \rightarrow 2$ edges indicate how the strengths can be modified in
    the case of the triangular lattice: in this case, one can set
    $d=\exp[-\beta w(e_d)]$ to account for the diagonal bond $e_d$
    perpendicular to the $2 \rightarrow 3$ edge. The Kasteleyn matrix
    has row $a$ and column $b$ indices $a,b=0,\ldots,5$.}
\end{figure}

Because each dimer configuration corresponds to a spin configuration
with the same energy, picking a sample from the dimer model with the
correct probability directly produces a corresponding spin
configuration that has the same probability of occurring.
We chose to
use Fisher cities for this work, instead of Kasteleyn cities
\cite{KastCities}, as they are simpler to sample using Wilson's
algorithm on a square lattice.  Also, by modifying the weights of the
Fisher cities, we can also very easily change the weights to simulate
triangular lattices (see Fig.\ \ref{fig:fish_city_mtx}).

\subsection{Matchings and the Kasteleyn Matrix}\label{SubSec:Kast}

Given the mapping between matchings using dual lattice cities and spin
configurations, we now briefly review the correspondence between dimer
matchings and Pfaffians.  Extensive discussion and examples can be
found in, for example, Refs.~\cite{KastCities,Robertson,KrauthBook}.
As a mathematical object, the Pfaffian $\Pf(A)$ can be
defined for general $2N\times 2N$ antisymmetric square matrices
$A=\{a_{qr}|q,r=0,\ldots,2N-1\}$, $a_{qr}=-a_{rq}$ by a restricted sum
over permutations $P=p(t)$ of the indices $t=0,1,\ldots,2N-1$,
\begin{equation}
  \Pf(A) = \sum_{P\,\mathrm{ordered}} (-1)^{\sigma(P)}a_{q_1r_1}a_{q_2r_2}\ldots
  a_{q_Nr_N}\ ,
\end{equation}
where $\sigma(P)$ is the sign of the permutation from the sequence $0,\ldots,2N-1$ to
the sequence $q_1,r_1,\ldots, q_N, r_N$ and the restriction to ordered $P$ is to
rearrangements where $q_k < r_k$, for all $1\le k \le N$, and $q_1 <
q_2 < \ldots < q_N$.  We also have that $[\Pf(A)]^2 = \mathrm{det}(A)$.

It turns out that summing over permutations with these two
restrictions is exactly the way to sum over dimer coverings for a
planar graph $G$, if the matrix elements of $A$ are chosen properly.
A matrix whose Pfaffian is $Z_{\mathcal M}=\sum_{M\in \mathcal{M}}
\exp[-\beta \mathcal{H}_G(M)]$ is the Kasteleyn matrix $K$.  This
matrix has entries $K(q,r)$, with $q,r=0\ldots,2N-1$, satisfying
$|K(q,r)|=x_{q,r}$, where $x_{q,r} = \exp\{-\beta w[e(q,r)]\}$, and
$w[e(q,r)]$ is the bond strength associated with edge $e(q,r)$.
Directions for the edges are then chosen so that all loops in $G$
which enclose an even number of nodes include an odd number of
counterclockwise edges \cite{KastCities}.  The matrix entry $K(q,r)$
is set to be $x_{q,r}$ if an edge is oriented from $q$ to $r$,
otherwise it is set to be $-x_{q,r}$.  This convention ensures that each
valid dimer configuration has positive net weight.  The Kasteleyn
matrix is thus a weighted version of a directed adjacency matrix.
Using these conventions and weight assignments gives \cite{KastCities}
\begin{equation}
  \Pf(K) = \sum_{M \in \mathcal{M}} \prod_{e\in M} x_e = Z_\mathcal{M} =
  Z^{-1}_R Z_\mathcal{J}\,.
\end{equation}
When decorating $G_0$ with cities to create $G$, the edges internal to the cities
must be assigned orientations.  An example of a Fisher city with the
correct directionality and the corresponding submatrix is shown in
Fig.\ \ref{fig:fish_city_mtx}.  The orientation of the connections
between the cities are from the 4-node in one Fisher city to the
0-node in the city to the right and from the 5-node a city to the
1-node in the city in the row above.  To simplify notation for the
rest of the paper, we will use $Z$ to indicate $Z_G$.

Established analytical and numerical techniques can be used to compute
$\Pf(K)=Z$.  As these numerical techniques require a number of
mathematical operations polynomial in the size of the lattice,
specifically growing as $\sim n^{3/2}$, the thermodynamic properties
can be efficiently computed. The number of bits needed for exact
computations grows with $n$, so that computing, for example, the exact
partition function, written out as a polynomial in $\exp(-\beta)$ of a
spin glass sample for the $\pm J$ model, where $J_{ij}=\pm 1$,
requires $O(n^{7/2})$ primitive fixed-word-length operations
\cite{GLV}.

We extend this correspondence to carry out sampling of spin
configurations by applying Wilson's algorithm.  Partial
diagonalization of the Kasteleyn matrix generates correlation
functions for the choice of the dimers in the matching representation.
These correlations are between dimers on a separator of the sample,
which divides the sample into two nearly equal places.  These
correlations functions include the probability of choosing any dimer
in a matching, so it is straightforward to determine whether a single
dimer is selected in a random matching.  The insight developed by
Wilson was to update these correlations as dimers are chosen: the
effects of partial assignment are propagated inductively to
correlations between other dimers, allowing many dimers to be assigned
without another factorization of the full Kasteleyn matrix. Once the
dimers have been selected on a separator, the two pieces are then
solved recursively, using their own separators.

\section{Wilson's algorithm}\label{Sec:Wilson}

In this section, we describe our implementation of Wilson's algorithm,
as applied and adapted to sampling configurations of the Ising spin
glass.  Wilson's algorithm samples dimer coverings: we map the Ising
problem to the dimer sampling problem using the mapping described for
the dual lattice in Sec.~\ref{Sec:Mapping}.  Wilson's algorithm uses a
``nested dissection'' \cite{LRTDissection}, i.e., a recursive
subdivision of the sample, where each subdivision of $n$ spins is into
two pieces of similar size separated by a line of vertices of size
$O(\sqrt{n})$, for efficiency.  Such a nested dissection was used by
Galluccio, Loebl, and Vondrak \cite{GLV}, to compute the full
expansion of the partition function of the $\pm J$ spin glass as a polynomial
in $\exp(-\beta)$, using the high temperature expansion formulation of
the partition function.  This dissection can be phrased using either a
dimer description, based on a matching of the decorated graph on the
dual lattice, or using spins. The algorithm is necessarily implemented
in terms of the former language, but for clarity, it is also
convenient to describe it using the latter language, i.e., based on
the spins on the original lattice.

Consider a subsample $U$ of Ising spins $\{s_i|i\in U\}$, possibly with
external fields at the boundary (corresponding to fixed spins
bordering $U$; this graph is still planar).  To divide this
sample into two independent samples, $U'$ and $U''$, a set $D$ of
spins is chosen as a spin separator, so that
\begin{equation}
  U=U'\cup D \cup U'' \label{Udissect}
\end{equation}
and no bonds connect spins in $U'$ to spins in $U''$.  We choose this
spin separator to be composed of two parallel lines of spins, so that
a line of nodes in the dual lattice is contained between the two
lines of spins.

It turns out that Wilson's approach provides an efficient way to
assign spin values along this separator, such that the spins are
selected with the correct probabilities.  That is, let such a spin
assignment on $D$ be $S_D=\{s_k=\pm 1| k \in D\}$. The spin at site $i$ for a
choice $S_D$ is also written as $S_D(i)$. One requires that the
probability that the algorithm will generate a particular choice $S_D$
is just equal to the probability $P_{\mathcal{J}}(S | S_D)$ that the
properly weighted choice of all spins will yield that particular
assignment of spins on the separator $D$, i.e., that
\begin{equation}
  P(S_D)=[Z(U)]^{-1}\sum_{S_U|S_U(i)=S_D(i), \forall i\in D}\exp[-\beta{\mathcal H}(S_U)]\,,
\end{equation}
with $S_U$ being a particular configuration of the spins in $U$, the
sum indexing all possible spin assignments consistent with the choice
$S_D$, and with $Z(U)=\sum_{S_U} \exp[-\beta {\mathcal H}(S_U)]$ the
partition function for $U$.  The remarkable property of the algorithm
to make such a selection implies that this procedure may then be
repeated on the remaining unassigned subsystems $U'$ and $U''$
independently of one another.

We can select the assignment for the spins in $D$ by sampling from the
dimer assignments for all the nodes in $\Delta$, where $\Delta$ is the
set of all nodes in $G$ that lie inside of $D$ and the connecting
edges contained within $D$.  This set of nodes $\Delta$ is what is
referred to as the separator in Wilson's work on an algorithm for
random dimer assignments.

In order to outline of our version of the algorithm for assigning
matchings in $\Delta$, one needs the notion of Pfaffian elimination
\cite{GLV}.  Let $K$ be a $2N\times 2N$ skew-symmetric matrix, i.e.,
$K(q,r)=-K(r,q)$ for $0 \le q,r < 2n-1$.  A cross operation between $q$
and $j$ is the addition of a multiple of row $q$ to row $r$ and the
same multiple of column $q$ to column $r$. If this multiple is given
by the factor $\alpha$, the cross operation on $K$ can be written as
\begin{equation}\label{eqn:alpha}
  K\rightarrow L(\alpha,q,r) K L^T(\alpha,q,r)\,,
\end{equation}
where $L(\alpha,q,r)$ is the lower triangular matrix $I + \alpha
\delta_{q,r}$. The matrix $\delta_{q,r}$ has all entries zero except
for a unit entry in row $q$ and column $r$. It turns out that the
value of $\Pf(K)$ is unchanged by cross operations, as $L$ has unit
determinant and $\Pf(B K B^T)=\det(B)\Pf(K)$ for general $B$
\cite{ElectronsMonteCarlo}.  Pfaffian elimination is the application
of multiple cross operations to simplify the matrix.  This
factorization via Pfaffian elimination has the goal of making the
Pfaffian trivial to compute; the simplest form of a skew-symmetric
matrix has non-zero values only in the even row superdiagonal
elements,
\begin{equation}\label{eqn_Sdef}
  Y=\sum_{\ell=0}^{N-1} y_{\ell}\sigma_2^{(\ell)}\,,
\end{equation}
where $\sigma_2^{(\ell)}$ is just the matrix that is non-zero except
for the $(2 \ell,2 \ell+1)$'st entry, which is set to $1$, and the $(2
\ell +1,2 \ell)$'st entry, which is set to $-1$.  In Pfaffian
elimination, then, the $\nu$ factors $\alpha_m$ and the cross operation
locations $q_m$, $r_m$ are all chosen sequentially so that
\begin{equation}
  Y = L K L^T\,,
\end{equation}
with $L=\prod_{m=1}^\nu L(\alpha_m,q_m,r_m)$ and $Y$ is of the form in
Eq.\ (\ref{eqn_Sdef}).  The needed choices of $\alpha_m$, $q_m$, and
$r_m$ are discussed in more detail in
Sec.\ \ref{sec:PfaffianFactorization}.

As the factorization of $K$ given by Pfaffian elimination leaves the
Pfaffian invariant
\begin{equation}
  \Pf(Y) = \Pf(L K L^T) = \det(L) \Pf(K) = \Pf(K)\ , \label{eqnPfFactor}
\end{equation}
the Pfaffian of the Kasteleyn matrix, and hence the partition
function, can be directly found by multiplying the even superdiagonal
entries of $Y$.

This elimination procedure resembles the application of Gaussian elimination to
compute the $LU$ factorization of a matrix $A$, with $A=LU$
where $L$ is lower triangular with unit elements on the
diagonal and $U$ is upper triangular.  The product of the diagonal
elements of $U$ gives $\det(A)$; here $\Pf(K)$ is the product of the
even row superdiagonal elements of $Y$.  Factorization via Pfaffian
elimination maintains the skew symmetry of the partially factorized
$\prod_m L(\alpha_m,q_m,r_m)\, K \prod_n L^T(\alpha_n,q_n,r_n)$ at each
stage.  Wilson presented his sampling algorithm using Gaussian
elimination; we find that Pfaffian elimination both clarifies the
algorithm and makes the programming of the algorithm more direct. A
version of the algorithm that we implemented using Gaussian
elimination was much less stable numerically than the one implemented
using Pfaffian elimination.

The factorization of $K$ given by Pfaffian elimination allows the
inverse of $K$ to be quickly computed. It is clear from
Eq.\ (\ref{eqnPfFactor}) that
\begin{equation}
  K^{-1} = L^T Y^{-1}L\ ,
\end{equation}
where, given the simple form of $Y$, the inverse of $Y$ is easily found:
\begin{equation}
  Y^{-1}= - \sum_{\ell=0}^{N-1} \frac{1}{y_\ell}\cdot\sigma_2^{(\ell)}\ .\label{eqnKinverse}
\end{equation}

\begin{figure*}
  \includegraphics[%
    width=\doublefigwidth,
    keepaspectratio]{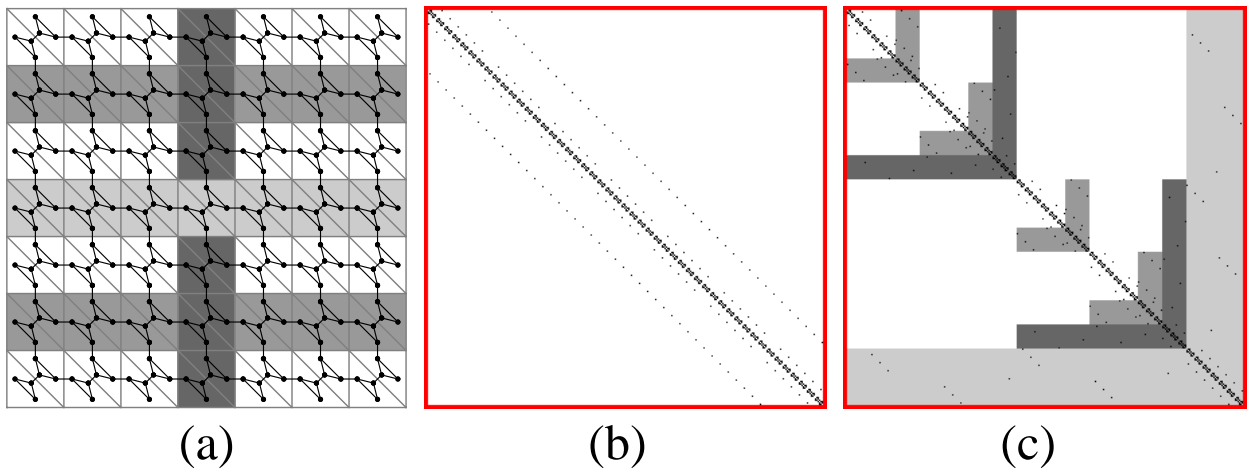}
  \caption{\label{fig:K_mtx}{[}color online{]} An example of the
    nested dissection and the Kasteleyn matrix $K$ for a $6\times 6$
    spin lattice sample surrounded by an outer layer of fixed
    spins. (a) The set of $8\times 8$ Ising spins sit on the sites of
    the light gray lattice of bonds of strength $J_{ij}$, where the
    diagonal bonds are indicated for the case of a triangular lattice.
    The graph $G$ on which the dimer sampling is computed is shown by
    the darker lines and circular nodes.  The gray bands indicate
    the nested dissection used for these nodes: the lighter gray region
    contains the dimer separator $\Delta\subset G$ and is bordered by
    the middle two rows of spins, the spin separator $D$. The darker and
    medium bands indicate, in order, the subsequent subdivisions of the
    sample.  (b) A display of the non-zero elements of the Kasteleyn
    matrix $K$, for a left-to-right and top-down ordering of the
    Kasteleyn cities.  The nonzero elements of the $294\times 294$
    Kasteleyn matrix for this are shown as black dots.  The edges
    internal to the Kasteleyn cities are closest to the diagonal:
    further non-zero elements represent connections between the
    cities.  (c) The permuted matrix $K$, where the cities are indexed
    according to a nested dissection.  The gray regions in $K$
    include connections contained within each of the separators of the
    nested dissection, with the same shades as in (a), and between the
    separators and other nodes.  The procedure of Pfaffian elimination
    can at most affect elements within the gray regions and also
    the values near the diagonal, for the nodes not contained in the
    gray regions in (a). Spin values are assigned to $D$ by
    examining the part of $K^{-1}$ indexed by the nodes of $\Delta$,
    i.e., the lower right square submatrix contained within the light gray
    region.}
\end{figure*}

 When the matrix $K$ is created, the indexing of the nodes in $G$ is
chosen according to a nested dissection of the graph $G$ that
maintains the grouping of the Kasteleyn cities. This ordering reduces
the amount of work needed to carry out the Pfaffian elimination and is
chosen so that the elements of the separator at each level of the
dissection are in a block at the lower right part of the submatrix
organized by that separator. An example of this ordering, given by the
nested dissection, is shown in Fig.\ \ref{fig:K_mtx}.

The core of the dimer assignment procedure is based on the
relationship between restricted partition functions and the Pfaffian
of submatrices of the Kasteleyn matrix. Consider two partition
functions, the entire partition function $Z=\Pf(K)$ and the restricted
partition function $Z_p$, which is sum of weights
$\prod_{e\in{G\setminus p}}x(e)$ restricted to matchings that include
the fixed partial matching $p=\{q_1,r_1,\ldots,q_k,r_k\}$, with
matched edges $(q_1,r_1),\ldots,(q_k,r_k)$. A listing of the terms
that contribute to $Z_p$ can be found by removing all nodes in $p$
from the graph $G$ and computing the Pfaffian of $K_p$, the Kasteleyn
matrix for $G\setminus p$. To find $Z_p$, the weights $x$ of the
removed edges must then be included, giving
\begin{equation}
Z_p = \Pf(K_p) \prod_{e\in p} x(e)\ .
\end{equation}
The weights $x(e)$ are uniform in Wilson's description, though he
noted the possibility of variable weights.
The probability $P(p)$ of choosing the edge set $p$ is therefore
\begin{equation}
P(p)=\frac{Z_p}{Z}= \frac{\Pf(K_p) \prod_{e\in p} x(e)}{\Pf(K)}\ .
\end{equation}
Given that one has already chosen an edge set $p$ that partially
covers a graph, the conditional probability $P(p,u\,|\,p)$ of edge $u$ being in
a complete matching that includes $p$ is
\begin{equation}\label{eqn:key}
P(p,u\,|\,p)=\frac{Z_{p,u}}{Z_p}=\frac{Z_{p,u}\cdot\Pf(K)}{Z_p\cdot\Pf(K)}
=\frac{\Pf(K_{p,u})x(u)}{\Pf(K_p)}\ .
\end{equation}

Fundamental relations between determinants and inverse matrices are
used in Wilson's algorithm to speed up the computation of $K_p$: we
directly adapt these relations for Pfaffian factorization.  Let $A$ be
a $2m\times 2m$ skew-symmetric matrix, and $0\le \ell < 2m$ be an even
integer, and $p=\{t_1,t_2,\ldots,t_\ell\}$ be a subset of indices for
the rows (columns) of $A$.  We will use the notation that
$A_p=A_{t_1,t_2,\ldots,t_\ell}$ denotes the $(2m-\ell)\times(2m-\ell)$
skew-symmetric matrix given by removing from $A$ all rows and columns
with indices in the set $(i_1,\ldots,i_\ell)$. The notation
$[A]_{t_1,\ldots,t_\ell}$ will denote the $\ell\times \ell$ matrix
resulting instead from keeping just those rows and columns and
eliminating the rest of the matrix.  Using this notation, and the
result that $\det(A)=[\Pf(A)]^2$, Jacobi's theorem (or directly using
the definition of the Pfaffian to show that element $i,j$ of $A^{-1}$
is $(-1)^{i+j}\Pf(A_{i,j})/\Pf(A)$) implies that \cite{Jacobi2}
\begin{equation}
  \frac{\Pf(A_{i_1,\ldots,i_\ell})}{\Pf(A)} = \pm \Pf([A^{-1}]_{i_1,\ldots,i_\ell})\,, \label{eqnJacobi}
\end{equation}
where the sign depends only on the choice
of the indices $i_1,\ldots,i_\ell$.

Eqs.\ (\ref{eqn:key}) and (\ref{eqnJacobi}) thus allow one to compute
the probability of matching $(q_1,r_1),\ldots,(q_k,r_k)$, using the
Pfaffian of the inverse of the Kasteleyn matrix where the same rows
and columns kept. The Pfaffian factorization of this latter matrix,
$\left[K^{-1}\right]_{q_1,r_1,\ldots,q_k,r_k}$, can be updated
incrementally as successive choices of matched edges are made. This
update allows for the progressive computation of the probabilities
$P(p,u\,|\,p)=x(u)z_k(u)$, where the updated factorization directly gives the value
$z_k(u)=\Pf\left(\left[K^{-1}\right]_{p,u}\right)/\Pf\left(\left[K^{-1}\right]_p\right)$.

Our adaptation of Wilson's algorithm can now be summarized in outline
form:
\begin{enumerate}
\item First, order the points of the decorated dual lattice $G$ in a
  manner consistent with the nested dissection.  The elements of the
  first dual separator $\Delta$ are at the end of this ordering.
\item Using this ordering, set the values of the Kasteleyn matrix $K$,
  which is stored as a sparse matrix.
\item Factorize $K$ using Pfaffian elimination. We use a pre-computed
  list of elementary operations to carry out the cross operations for all
  elements that are potentially non-zero.
  (Stop here if only the partition function for $U$ is
  needed; the partition function $Z$ is just the product of alternate
  superdiagonal elements in $Y$, i.e., $Z=\prod_{\ell=0}^{N-1} y_\ell$.)
\item\label{step4} Using this factorization, compute the elements of $K^{-1}$ that
  are indexed by elements of $\Delta$; this is $[K^{-1}]_\Delta$, the
  lower-right submatrix of $K^{-1}$ with indices contained in
  $\Delta$. (For some speed up, as suggested in Ref.\ \onlinecite{Wilson},
  we only compute the elements of $\left[K^{-1}\right]_\Delta$ that are needed
  in the following steps, at the time those elements are required.)
\item Assign dimers $(q_1,r_1),(q_2,r_2),\ldots$ along the separator $\Delta$:  
  \begin{enumerate}
  \item \label{updatestep}
    Choose a node $q_1\in \Delta$ such all edges that are incident upon $q_1$ are
    fully contained in $\Delta$.
    Choose among the potential edges $(q_1,r_1)=e$ with the probabilities
    $K(q_1,r_1)\Pf({K}_{q_1,r_1})/\Pf(K)=K(q_1,r_1)\Pf([K^{-1}]_{q_1,r_1})$.
  \item Repeat this last substep, \ref{updatestep}, proceeding along the dimer and updating
    $[K^{-1}]_{q_1,r_1,\ldots,q_k,r_k}$ and its factorization, until no more matchings
    can be added wholly within the set $\Delta$.
  \end{enumerate}
\item Use the dimer matching for $\Delta$ to assign spin values in the
  spin separator $D$, which surround the dimer separator $\Delta$.
\item Recursively repeat items 1-7 for the subproblems $U'$ and $U''$.
\end{enumerate}

Note that, in some cases, an alteration of this procedure can be used
to speed up this method. It might be that faster results can be
obtained using simple floating point numbers (double precision),
rather than multi-precision numbers, though they may not provide
numerical accuracy to carry out all of the calculation. A compromise
would be to carry out the computation for only part of the separator
at a time, making the computation more stable.  The whole matrix $K$
with the remaining unchosen nodes is recomputed and the process is
repeated.  This method is asymptotically slower, but practical
for systems of intermediate size at intermediate temperature.

\subsection{Entries of $K$: nested dissection and storage}

The Kasteleyn matrix $K$, as defined in Section \ref{SubSec:Kast}, is
indexed by the nodes of the decorated dual graph $G$.  As the entries
$K(q,r)=\pm x(q,r)$ of $K$ are non-zero only for entries indexed by neighboring
points $q$ and $r$ on the decorated dual lattice, this $O(n) \times
O(n)$ matrix has only $O(n)$ non-zero entries.  If the nodes are
indexed in a natural, geometric, lattice order, the Kasteleyn matrix
$K$ is simple, as shown in Fig.\ \ref{fig:K_mtx}(b).
However, matrix manipulations, such as Pfaffian elimination, for general
matrices might lead to the computation of $O(n^2)$ non-zero entries.

To compute the correlations between spins on the separator, the nodes are reordered,
though kept together in city groups. In this reordering, the nodes are
each assigned a new index. This reordering satisfies the nested
dissection property that, at each level, the separator nodes in
$\Delta$, which give the spin sub-sample $U$, have the highest
index. This implies that the non-zero values defined by the weights
contained within the separator $\Delta$ are at the lower and rightmost
parts of $K$, at each level, while the non-zero values for nodes
belonging to $U'$ and $U''$ [see Eq.\ (\ref{Udissect})] are confined
to square blocks about the diagonal. An example of the distribution of
matrix entries, given this ordering of the nodes $V$ of $G$, is shown
in Fig.~\ref{fig:K_mtx}(c). This organization confines all matrix
manipulation to a portion of the shaded regions of the matrix and to a
narrow band around the diagonal, as unshaded entries away from the
diagonal always have value zero. The shaded regions make up
$O(N^{3/2})$ entries, though only a subset of even those entries,
growing with $N$ approximately as $\sim N$, possibly with a
logarithmic correction, are used in the Pfaffian elimination.

Given our specific choice of separator, the nodes of $G$ corresponding to the
Kasteleyn cities always form subsequences in the ordering of the nodes. That is,
they remain grouped together. Note that the
submatrices for each city are uniform in structure. This choice of separator
$\Delta$ (as all of the dual nodes between two rows or columns of spins) is not
the most efficient, as slightly smaller separators $\Delta \subset G$ can be
chosen, but it is a very convenient choice that maintains a uniform structure.

We use this ordering to construct $K$ as a sparse matrix, using $O(N)$
operations and time.  The sparse matrix storage scheme is relatively
direct (see, e.g.\ \cite{SparseMatrixRef} for a discussion on sparse
matrix algorithms and storage techniques). We have the advantage here
that, for the Pfaffian elimination, both the locations of the needed
elements and the list of operations using these elements can be
pre-computed and stored on disk.  This allows us to place the elements of the matrix $K$
in a linear array with $O(N)$ elements, with the elements ordered by
the step at which they are first needed in the Pfaffian
elimination. This precomputation is independent of both the data type
that we use and the bond strengths for the spin lattice.

\subsection{Pfaffian factorization}\label{sec:PfaffianFactorization}

\begin{figure}  \includegraphics[%
    width=\singfigwidth,
    keepaspectratio]{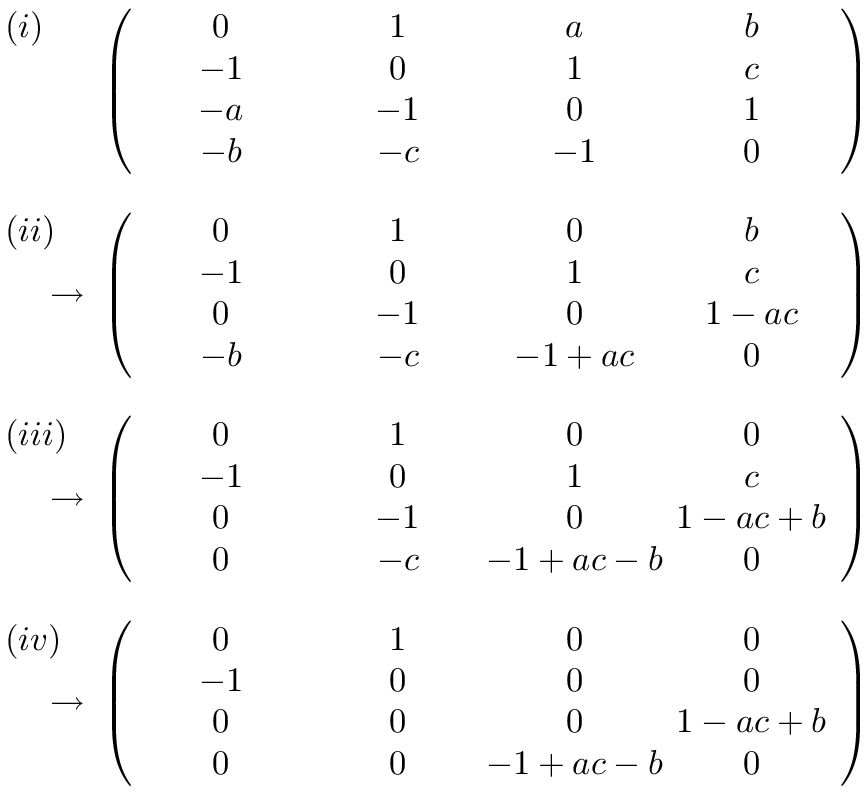}
  \caption{\label{fig:cross} Example of cross operations used for
    Pfaffian elimination.  (i) A skew-symmetric matrix $K$. (ii) The
    result of a cross operation $K\rightarrow
    L(\alpha,i,j)KL^T(\alpha,q,r)$ of the first type, with $q=0$,
    $r=2$, $\alpha=-a$, applied to $K$. This is found by adding
    $\alpha=-K(q,r)/K(q,q+1)$ times column $q+1$ to column $r$ and
    then $\alpha$ times row $q+1$ to row $r$, to eliminate the element
    at location $(q,r)$.  (iii) The result of the next cross
    operation, with $q=0$, $r=3$, and $\alpha=-b$. (iv) The result of
    two subsequent operations of the second type, where
    $\alpha(q,r)=-K(q,r)/K(q,q-1)$, for $q=1$, $r=2$ and $q=1$,
    $r=3$. These latter types of operation are not needed to compute
    $\Pf(K)$, but are needed for finding $[K^{-1}]_\Delta$.
    The Pfaffian of $K$ is the product of the
    superdiagonal elements in even rows: here, $\Pf(K)=(1)(1-ac+b)$.  }
\end{figure}

Pfaffian elimination and the concomitant factorization of $K$ proceed
by the elimination of elements by cross operations. There are two
types of cross operations that are carried out.  The first type of
operation eliminates all but the first of the non-zero entries in an
even row.  This is done for an even row $q$ by using (see Eq.\ \ref{eqn:alpha})
$\alpha(q,r)=-K(q,r)/K(q,q+1)$ for all $r \ge q+2$.  The second type
eliminates all entries in odd rows. This is done for odd $i$ using
$\alpha(q,r)=-K(q,r)/K(q-1,q)$ with $r\ge q$. Examples of
operations of each type are traced out in Fig.\ \ref{fig:cross}.

We note that in carrying out Pfaffian elimination, a potential danger
would be that one of the even-row superdiagonal elements, $K(q,q+1)$
with $q$ even, is zero. In this situation, it would be necessary to do
a pivoting operation, which would destroy the nested dissection.
However, given that the Kasteleyn cities remain grouped together,
the sequential pairing of nodes
$(0,1),(2,3),\ldots$ is always a matching. Hence the Pfaffian of any upper left portion of
the Kasteleyn matrix, as we have arranged it, is non-zero, as the
Pfaffian counts matchings (in a weighted fashion), and there is always
a matching for the upper left portion of the matrix of unit weight.
This implies that all superdiagonal elements in the even rows must be
non-zero.  This provides a ``built-in'' version of the permutation of
nodes to accommodate a matching that is given in Wilson's paper
\cite{Wilson}. In the periodic case (Sec.\ \ref{sec:torus}), for
certain boundary weight choices at $\beta=0$ ($T=\infty$), when the
bond strengths have uniform magnitude, there can be ``accidental''
cancellations which will cause this procedure to fail, as the signed
weight of a sub-matching can be exactly zero, even though the Pfaffian is
non-zero. In this case, permutation of the remaining elements of the
matrix (i.e., ``pivoting'') would be needed to remove a zero from the
superdiagonal and obtain the correct factorization.

The factorization found by Pfaffian elimination, Eq.\ (\ref{eqnPfFactor}),
then allows for the easy computation of the partition
function for the given sample, at the temperature used to set the elements of
$K$, if desired. The Pfaffian of the original Kasteleyn matrix is simply the
product of the even superdiagonal elements of $Y$,
\begin{equation}
  \Pf(K)=\prod_{\ell=0}^{N-1} y_{\ell}\ .
\end{equation}
Note that this is the procedure, computation of the Pfaffian of $K$
using nested dissection, was used by Galluccio {\em et al.} \cite{GLV}
to compute the partition function. In that work, to compute the
partition function at a given temperature, the arithmetic is carried
out modulo prime integers, for a selection of prime integers. The
partition function at that temperature is then reconstructed by
application of the Chinese remainder theorem. The whole partition
function as a function of $\beta$ can be found by polynomial
interpolation in $\exp(-\beta)$. This full calculation works only if
the couplings $J_{ij}$ are restricted to small integer values,
typically $J_{ij} = \pm 1$.

\subsection{Sampling: inductively factorizing $K^{-1}$}

At this point, though one has the partition function (from the even
superdiagonal elements of $Y$), sampling spin configurations requires
a bit more work.  The sampling can be carried out by using only the
lower right hand corner $[K^{-1}]_\Delta$ of $K^{-1}$.  This part of the
matrix encodes all the correlations between the spins in $D$, on the
separator of the sample, via the correlations of dimer coverings of
$\Delta$.  These correlations are used to make dimer (and then spin)
assignments along the geometric separator. The description in this subsection
is based upon Wilson's description and notation \cite{Wilson}, only with
a change in the factorization method (Pfaffian vs.\ Gaussian).

To assign a dimer covering inside the separator $\Delta$, the
algorithm proceeds through each of the edges in $G$ that are wholly
contained within the node set $\Delta$ and computes the probability
that that edge is covered by a dimer, conditioned on earlier
assignments of dimers in the separator. The algorithm proceeds
inductively by calculating the probabilities for placing the $(k+1)$'st
dimer using the results of the calculations for the previous $k$ edges
in $\Delta$, $p=\{(q_1,r_1),\ldots,(q_k,r_k)\}$.

The inductive computation of the probabilities are based on
Eq.\ (\ref{eqn:key}), which in turn requires the computation of the
ratio $\frac{\Pf(K_{p,u})}{\Pf(K_p)}$.  This ratio is found from
the change in the Pfaffian of
$[K^{-1}]_{q_1,r_1,\ldots,q_k,r_k}$ that results from the augmentation of $[K^{-1}]$
by two rows and columns, those with indices $q_{k+1}$ and $r_{k+1}$ in $K^{-1}$. To calculate this
change, the algorithm maintains a factorization of
$A_k\equiv[K^{-1}]_p$ which is tentatively updated to test the
addition of an edge. This factorization allows for the ratios of Pfaffians
to be quickly computed.
The matrix $[K^{-1}]_\Delta$ is first found by computing a subset of
the rows and columns of $K^{-1}$ using Eq.\ (\ref{eqnKinverse}) and
the Pfaffian factorization of $K$, Eq.\ (\ref{eqnPfFactor}).

To select matched edges within $\Delta$, one considers in turn nodes
$q\in\Delta$ such that all neighbors $r$ of $q$ are also in $\Delta$
and selects one of these neighbors with the correct probability.
When considering matches for such a node $q_{k+1}$, assume that one
has already selected $k$ dimers in $\Delta$, as part of a sampling
inside $\Delta$, and that one knows the matrices $M_k$ and $V_k$ in
the factorization
\begin{eqnarray}
  M_k A_k M_k^T = V_k,\ \label{eqnAfactorize}
\end{eqnarray}
where all matrices in this equation are of dimension $2k\times 2k$,
$M_k$ is lower triangular, and $V_k$ has the same super-diagonal
structure as $Y$.  For a given trial edge $(q_{k+1},r_{k+1})$, we can
tentatively extend the matrices $M_k$ and $V_k$ to $M_{k+1}$ and
$V_{k+1}$, with
\begin{equation}
  M_{k+1} = \left[\begin{array}{cc}
      M_k & 0 \\
      m_{k+1} & I
    \end{array} \right]
\end{equation}
and
\begin{equation}
  V_{k+1} = \left[\begin{array}{cc}
      V_k & 0 \\
      0   & v_{k+1}
    \end{array} \right]\,,
\end{equation}
where $v_k+1$ is a $2\times 2$ antisymmetric matrix,
\begin{equation}
v_k = \left[\begin{array}{cc}0 & z_{k+1}\\ -z_{k+1} & 0\end{array}\right]
\end{equation}
and $m_{k+1}$ is a $2\times 2k$ matrix.  To compute these trial
solutions $M_{k+1}$ and $V_{k+1}$, one first tentatively updates
$A_{k+1}$,
\begin{equation}
  A_{k+1} = \left[\begin{array}{cc}
      A_k & -a_{k+1}^T \\
      a_{k+1} & b_{k+1}
    \end{array} \right]\,,
\end{equation}
using the rows and columns indexed by $q_{k+1}$ and $r_{k+1}$ from
$[K^{-1}]_\Delta$ to fill in $A_{k+1}$ and reading off $a_{k+1}$ and
$b_{k+1}$.  Direct matrix multiplication and requiring
Eq.\ (\ref{eqnAfactorize}) for $A_{k+1}$ then give
\begin{equation}
  m_{k+1} = -a_{k+1}A_k^{-1} = - a_{k+1} M_k^T V_k^{-1} M_k\label{eqnMupdate}
\end{equation}
and that
\begin{equation}
  z_{k+1}=b_{k+1} + a_{k+1} M_{k}^T V_k^{-1} M_k a_{k+1}^T\ .\label{eqnz}
\end{equation}

\begin{figure}
  \includegraphics[%
    width=\singfigwidth,
    keepaspectratio]{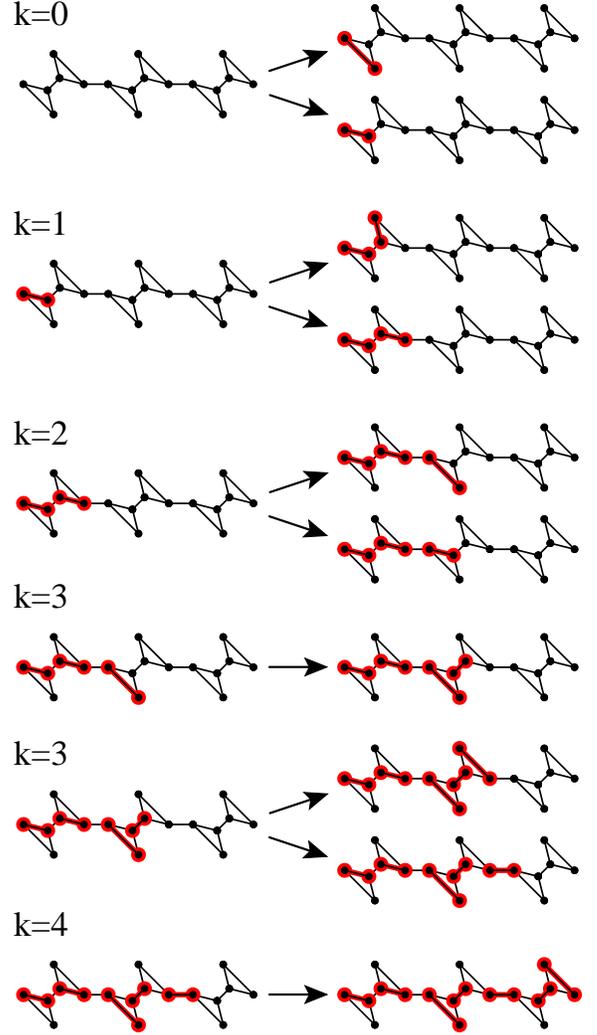}
  \caption{\label{fig:assigning_dimers} {[}color online{]} An example
    of the dimer assignment procedure for a separator $\Delta$ that is
    three cities wide. Initially, no edges are matched (left part of $k=0$).
    The first choice, $k=0$, is between
    the two edges inside $\Delta$ that are incident upon the far left node.
    In the example shown, the lower bond
    (connecting node 0 to node 2; see Fig.\ \ref{fig:fish_city_mtx})
    is chosen, as shown on the left of the $k=1$ section of the
    figure. At this stage, one has computed matrices $A_1$, $M_1$, and
    $V_1$. The comparison of the next two possible matchings, shown on
    the right part of the $k=1$ subfigure, compares the inclusion of the
    $(q,r)=(3,5)$ and $(3,4)$ edges. In some cases, as in the first
    $k=3$ panel, a choice is forced and $A_k$, $M_k$, and $V_k$ need
    not be updated. The $k=4$ choices are forced, as an even number of
    domain walls must cross the separator, so that an even number of
    the top nodes and an even number of the bottom nodes are
    unmatched.}
\end{figure}

As $\Pf(A_k)=\prod_{i=1,..,k} z_i$, the factor $z_{k+1}$ is the ratio
$\Pf(A_{k+1})/\Pf(A_k)$ of the Pfaffians that is needed to apply
Eq.\ (\ref{eqn:key}). Hence, this update in the factorization allows
us to find the probability
$x_{q_{k+1},r_{k+1}}z_{k+1}(q_{k+1},r_{k+1})$ of selecting the
specific edge $(q_{k+1},r_{k+1})$ to augment the matching.  Once we
have chosen a match for $q_{k+1}$, we then update $A_{k}$ to $A_{k+1}$
from $K^{-1}$, $V_k$ using $z_{k+1}$, and $M_{k}$ using
Eq.\ (\ref{eqnMupdate}).  This process is repeated until a maximal
(though usually not complete) matching within $\Delta$ is
obtained. With our choice of Fisher cities, there are only two
candidates $r_{k+1}$ for each $q_{k+1}$ when using fixed boundary
conditions; for periodic boundary conditions (Sec.\ \ref{sec:torus}),
matching the initial node $q_1=0$ requires the comparison of three
choices.  Note that not all the $z$ need be computed as the total
probability sums to unity; when considering two choices, considerable
time is saved by computing the probability of only one of the choices.
An example of dimer assignment is depicted in
Fig.\ \ref{fig:assigning_dimers}.

The results derived by Wilson for the bounds on the number of steps
using Gaussian elimination carry over directly to the approach using
Pfaffian elimination. The maximal size of the separator is of order
$O(L)=O(n^{1/2})$.  There are $O(n^{3/2})$ operations in the dimer
assignment for the largest separator: matching a single dimer requires
at most $O(n)$ steps, due to the multiplication of matrices of size
$2\times O(n^{1/2})$ by matrices of size $O(n^{1/2})\times
O(n^{1/2})$, and there are $O(n^{1/2})$ matchings in each
separator. Calculating $K^{-1}$ also requires $O(n^{3/2})$ steps.  As
the smaller separators decrease in size geometrically, as the sample
is subdivided, the number of operations for each of the smaller
separators decreases geometrically, and the sum of steps over all
levels of the nested dissection gives a total of $O(n^{3/2})$
arithmetic steps to generate a random assignment.  The running time
then is a product of the time per operation, which depends on the
needed precision, and this number of steps. As discussed in more
detail in Sec.\ \ref{subsec:typesprecision}, the running time grows
roughly linearly with the precision: the necessary precision grows only slowly
with $n$, but proportionally to $\beta$.

\begin{figure}
  \includegraphics[%
    width=\singfigwidth,
    keepaspectratio]{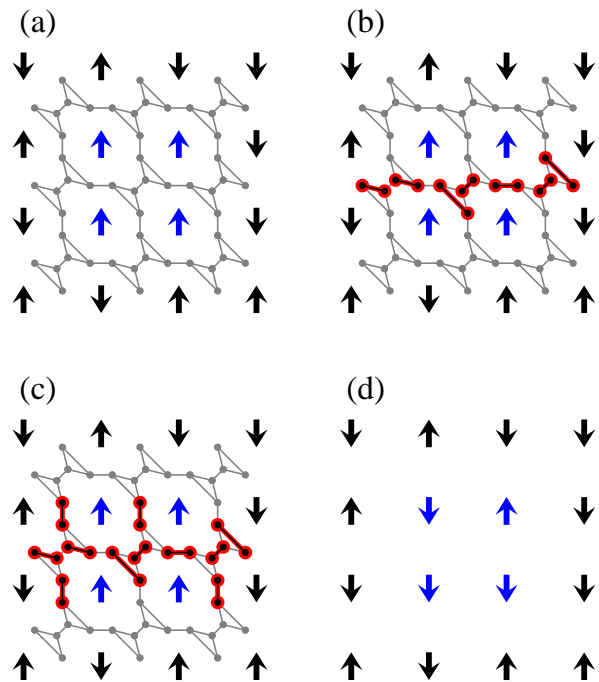}
  \caption{\label{fig:assigning_spins} {[}color online{]} Application
    of the result of the example dimer assignment from
    Fig. \ref{fig:assigning_dimers} to the spin assignment.  (a) The
    initial spin configuration, with fixed spins on the boundary.
    The portion of
    $G$ used to compute $K$, $[K^{-1}]_\Delta$ and the dimer assignment
    is indicated in gray. The middle row of Fisher cities composes
    $\Delta$, the dimer separator. (b) The sample dimer assignment
    (partial matching) for $\Delta$ from
    Fig.\ \ref{fig:assigning_dimers} superimposed on $G$. (c) Extra
    choices in the matching are forced by the matching internal to
    $\Delta$.  These additional dimers cut across the bonds
    separating spins in $D$, the two spin rows parallel to $\Delta$.
    (d) In the last step, the modifiable spins are updated. The update
    is based upon the portion of domain walls forced by the partial
    matching in (c).  Moving from left to right, for example, from the
    two fixed spins on the middle of the left side, a spin is reversed
    if an odd number of dimers extending from $\Delta$ are crossed.  }
\end{figure}

Once all nodes in the separator $\Delta$ have dimers associated with
them, the broken bonds along the strip $D$ of the Ising system are
found from the locations of the dimers between these cities and the
neighboring ones.  We can then directly assign the spins along the
strip. An example of such a spin assignment is displayed in
Fig.\ \ref{fig:assigning_spins}.

\subsection{Verification}

The structure of the calculation is rather complex, so we verified our
implementation of the algorithm in several ways. We checked exact
partition function calculations for pure systems against the results
of our computation. Exact enumeration for pure and disordered samples
in systems up to $n=5^2$ was used to predict sampling probabilities:
we then used our code to generate over $10^5$ samples and compared the
sampled probability distribution with the exact calculations. These
were in statistical agreement. Each author of this paper developed a
code independently: these were compared on the same Gaussian spin
glass samples of size $33^2$ and found to generate the same
distribution for configurations, at low temperatures, also consistent
with the Boltzmann distribution for total energy.  At low
temperatures, the sampled configurations approached those of the
ground state configurations (which were predicted using an independent
ground state code based on combinatorial optimization methods
\cite{Barahona,GSKastCities}).

\subsection{Data types and timing}\label{subsec:typesprecision}

Our code is constructed so that the data type of matrix elements can
be any field (double precision numbers, multiple-precision numbers, or
exact rationals, for example). This allows us to check the effects of
the choice of numerical type on the accuracy, stability, and running
time of the sampling algorithm.  For higher precision variables, we
use the GMP library \cite{GMP} for exact rational arithmetic and
either the MPFR \cite{MPFR} extension to GMP or the GMP library itself
for multiple-precision floating point arithmetic.  We find that the
latter two floating point types give comparable performance and
accuracy. Using exact rationals allows for mathematically exact
sampling, but results in a temperature-dependent slowdown by a factor
of $10$ or $100$ over the range of temperatures, $T=0.1$ to $T=1$, we
used while comparing rationals with floating point calculations.

\begin{figure}
  \includegraphics[%
    height=\singfigwidth,
    angle=270,
    keepaspectratio]{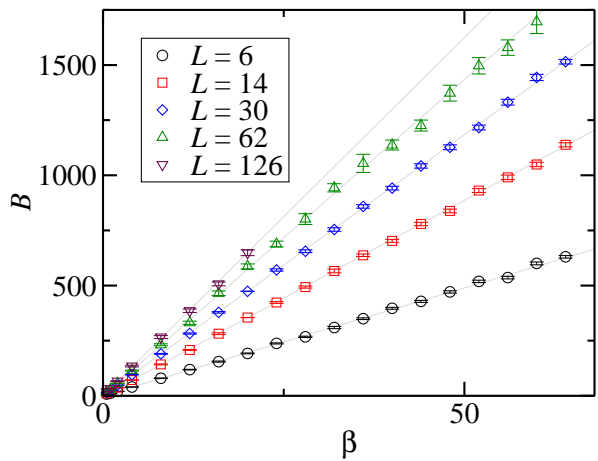}
  \caption{\label{fig:failbits} {[}color online{]} The plot shows the
    sample averages of the number of bits of precision $B_{\mathrm
      avg}$ required to obtain the correct sampling, for system sizes
    $n=L^2=6^2$ through $126^2$ with fixed boundaries, as a function
    of inverse temperature $\beta$, for Gaussian disorder. The lines
    indicate linear fits of the form $B_{\mathrm avg}=c\beta$. The
    number of bits needed for periodic boundary conditions (not shown)
    are very close to these same lines, at each system size. To find
    an accurate result with high confidence, one can use twice the
    average needed value: this was sufficient for all samples
    ($>10^4$) that we examined.}
\end{figure}

The edges $u$ are chosen by comparing the probability $P(p,u\,|\,p)$ with
a random number chosen in the interval $\left[0,1\right)$. The
sequence of random numbers and computed probabilities determines the
spin configuration selected.  We determine the needed precision for a
given sample and temperature by demanding that the result of a
specific assignment be independent of the precision, for a given
sequence of random numbers.  Note that using this precision does not
give the exact values of the probabilities at each stage of the
computation, but the sampling does not change at increased precision.
If a number in the sequence happens to be extremely close to the
computed probability, higher precision arithmetic could be required.

Results of our tests for needed precisions are summarized in
Fig.~\ref{fig:failbits}, where we plot the number of bits needed,
determined by bisection in the number of bits, averaged over
random number sequences and disorder. We find that the distribution of the required number of
bits is not very broad, regardless of temperature and disorder
realization $\mathcal{J}$. Less than $10^{-4}$ of the attempts require
more than double the average precision to find the correct sampling.
Hence fixing the precision at two to three times the average value
will almost guarantee an exact sampling.

\begin{figure}
  \includegraphics[%
    height=\singfigwidth,
    keepaspectratio]{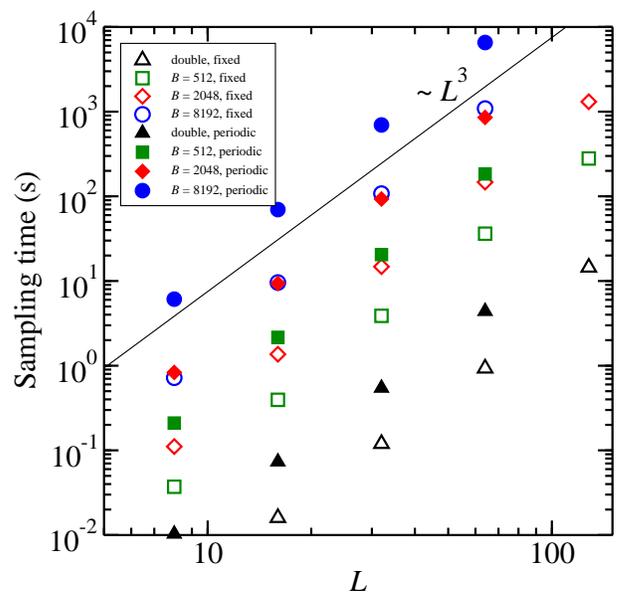}
  \caption{\label{fig:timings} {[}color online{]} Run time, measured
    in seconds, to generate a single configuration, as a function of
    system size $L$, using a 2.4 GHz Intel Core 2 Duo processor (MacBook Pro).
    Double precision (64 bit floating point) data is
    indicated with triangles, while multi-precision results for
    $B=512$, 2048, and 8192 bits are indicated by squares, diamonds,
    and circles, respectively.  Samples are generated for $L\leq128$
    with fixed boundaries (closed symbols) and for $L\leq64$ with periodic
    boundaries (filled
    symbols). The sample-to-sample fluctuation of
    disorder realization is less than 0.1\% of the run time, so error
    bars are not shown. The solid line indicates the form of the
    expected dependence of run time on system size for a given fixed
    precision, that is, $\sim L^3$.}
\end{figure}

For high temperatures (of order $T = 1$), low
precisions (i.e.\ fixed double precision variables) are sufficient for the
system sizes we study (see Fig.\ \ref{fig:timings}).  For lower temperatures,
higher precisions are needed. The needed precision is well fit by a linear
growth in $\beta$, for $\beta>0.5$. This is consistent with the expectation
that, as the weights vary as $\exp(-\beta J)$, the number of bits needed
to describe the weights grows linearly with $\beta$, for fixed typical values
of $J$. The number of needed bits grows only slowly with $L$. This is consistent
with the structure of the sampling and Pfaffian computation, which are hierarchical
in structure, so that the accumulated error grows only slowly with $L$.

For systems up to size $64^2$, $600$ bits of precision are sufficient
for temperatures $T>0.1$. For larger systems and lower temperatures,
more bits are needed. For example, we use 2048 bits to reliably
sample configurations at $\beta=25$ and $L=128$.

We collected timing data for the performance of our algorithm as a
function of system size and temperature. These data are summarized in
Fig.\ \ref{fig:timings}.  We find that sampling with periodic boundary
conditions (Sec.\ \ref{sec:torus}) takes approximately 5.5-6.5 times
longer than sampling with fixed boundary conditions.
The needed precision and running times for $\pm J$ disorder are very close
to those shown in Figs.\ \ref{fig:timings} and \ref{fig:failbits}.
For $64<B<512$,
the run time to sample a configuration varies only slowly with $B$,
approximately by a factor of $1.5$ over this range. For higher precision,
the running time grows somewhat faster than linearly with $B$, and
hence somewhat faster than linearly with $\beta$.

\section{Periodic Boundary Conditions}
\label{sec:torus}

Fixed boundary conditions are appropriate for patchwork dynamics, but,
for other simulations, other boundary conditions, may be
useful.  One simple way to implement open boundary conditions is to set to zero all the $J_{ij}$ connecting interior to
boundary spins.  For
cylindrical samples with open boundaries, we use a ``separator''
which does not actually separate the graph, but one that slices the
sample perpendicular to the circumference of the cylinder, resulting
in a simple planar graph with fixed boundaries.  Toroidal graphs
require a more complicated sampling scheme, as they are not planar.
In general, for a graph of genus $g$, the partition function of the
dimer problem may be calculated exactly by summing $4^g$ Pfaffians
\cite{KastTorus}.  The reasoning behind this summation can be adapted
to sampling for periodic spin lattices.

\subsection{Partition function on the periodic lattice}

The Kasteleyn matrix approach for computing $Z$ can be extended to
handle the periodic case, by adding connections between cities that
complete the periodic boundaries, converting the planar square sample
to a toroidal one, but the direct correspondence between dimer
configurations and spin configurations is affected.  On the torus,
topologically non-trivial domain walls must always come in pairs, or
the spin configuration can not be consistently defined. But the
matching problem allows for odd numbers of loops to wrap around the
torus on either axis. For $T=0$ ground states, one can decide to
ignore this fact and allow variable boundary conditions, which allow
for an odd number of domain walls relative to other boundary
conditions. Choosing the boundary condition and spin configuration
that jointly minimize $\mathcal{H}$ gives the extended ground state
construction \cite{KastCities}.  At finite temperature with fixed
boundary conditions, however, we need to arrange for the cancellation
of dimer configurations which would imply an odd number of domain
walls that wrap around either axis.

\begin{table}[tb]
  \centering 
  \begin{tabular}{c|cccc}
    & $\Pf(K^{++})$
    & $\Pf(K^{+-})$
    & $\Pf(K^{-+})$
    & $\Pf(K^{--})$ \\ \hline
    (e,e) & + & + & + & + \\
    (o,e) & - & - & + & + \\
    (e,o) & - & + & - & + \\
    (o,o) & - & + & + & -
  \end{tabular}
  \caption{\label{cap:pfaff_table} A table of the signs for different
    combinations of spanning loop parities in the dimer model for each
    of the four Pfaffians $K^{\pm\pm}$ for the torus. The set of loops
    found from a dimer configuration can have a total wrapping number
    that is odd (o) or even (e) number along either the horizontal or
    vertical directions. This gives four possible classes of dimer
    configurations (e,e), (e,o), (o,e) and (o,o).  For the dual
    mapping used here, the physical spin configurations for the Ising
    model are restricted to those with an even number of domain walls
    wrapping in both directions, i.e., the (e,e) class.  The four
    classes of dimer configurations are summed in each Pfaffian of the
    four Kasteleyn matrices, $K^{\pm\pm}$, with a sign that depends on
    the class and the matrix.  These four matrices assign different
    signs to the weights of the dual edges that connect the boundaries
    together, with a $+$ or $-$ sign for each of the two types of
    boundary connections, i.e., horizontal or vertical.  Applying this
    table, we get the partition function for the valid dimer
    configurations by the sum
    $Z=[\Pf(K^{++})+\Pf(K^{+-})+\Pf(K^{-+})+\Pf(K^{--})]/2$, which counts only the
    (e,e) class of dimer configurations. This sum differs from the
    more commonly studied case, the dimer model using cities on the
    primal lattice, where all classes of matchings are valid
    configurations and $Z=[-\Pf(K^{++})+\Pf(K^{+-})+\Pf(K^{-+})+\Pf(K^{--})]/2$
    gives the sum over (e,e), (o,e), (e,o), and (e,e).}
\end{table}

This cancellation is achieved by summing over four Pfaffians, in a fashion similar to
that developed for the primal lattice \cite{KastCities}, though the
details differ for the dual lattice. The four Pfaffians correspond to
four possible choices of sign for the elements of $K(q,r)$ that
complete the periodic connections.  That is, the values of $K(q,r)$
for edges that connect the last column to the first column (that wrap
around in the $x$ direction) are uniformly set to one of two choices,
$\pm\exp[-\beta w(q,r)]$, and the values for the edges that connect
the last row to the first row (that wrap around in the $y$ direction)
are also uniformly set, independent of the choice for the $x$-wrapping
bonds, again to $\pm\exp[-\beta w(q,r)]$. This
gives four matrices, $K^{++}$, $K^{-+}$, $K^{+-}$, and $K^{--}$. The
dimer configurations that are summed up in the Pfaffians enter with
different relative signs, depending on how many times the matchings
wrap around each axis, as the parity of the windings affects the sign
of the dimer configurations when the negative sign is chosen for the
periodically-connecting edges. The effects of these signs are
tabulated and explained in Table\ \ref{cap:pfaff_table}. The sum of
the Pfaffian of these four matrices then gives twice the partition
function, as those dimer configurations with an even number of
wrapping loops enter four times and those with an odd number, in
either direction, are cancelled out, and there is a two-to-one mapping
of spin configurations to domain walls in the periodic case (due to
global spin flip symmetry).

\subsection{Matching probabilities for the torus}

There are several simple possible choices for a nested dissection for
toroidal samples of dimension $L\times L$.  The number of cities is
the same as the number of variable spins, i.e., $L^2$.  We chose to
use a horizontal strip of length $L$ in the first row of cities, which
fixes the spins in the first two rows, followed by a vertical strip in
of length $L-1$ in the first column, which fixes the first two columns
of spins, followed by a sampling the remaining $(L-1)\times(L-1)$
cities, i.e., a sampling of the remaining $(L-2)\times(L-2)$ spins
using the already determined spins in the first two columns and rows
as fixed spin boundary conditions. The first two ``separators'' don't
divide the sample into separate pieces, but instead provide for the
cutting of loops that wind around the torus, in two stages.

For the first sampling, the periodic horizontal row, one has to sample
using four Kasteleyn matrices in parallel. For the second sampling, on
a cylindrical geometry, one needs to find probabilities by summing
over two Kasteleyn matrices $K^+$ and $K^-$, in order to eliminate
domain walls that wrap around the cylinder an odd number of times. We
can consider both cases as specific examples of a general problem:
sampling using multiple Kasteleyn matrices simultaneously.

For this general case, consider a partition function $Z$ that is found
by summing the Pfaffian over matrices $K^{\alpha}$, with weights
$q_\alpha$ (e.g., $\alpha={\pm\pm}$ and $q_\alpha=\frac{1}{2}$ for
toroidal boundary conditions). The partition function is then
\begin{equation}
Z=\sum_\alpha q_\alpha K^\alpha\ .
\end{equation}
The computation of probability of selection is more complicated than
for the case of a single $K$.  For each $K^{\alpha}$, we consider the
inverse indexed by elements of the separator $\Delta$,
$[(K^\alpha)^{-1}]_\Delta$, and inductively factorize
$[(K^\alpha)^{-1}]_p$ for our current choice of sampled edges
$p=\{e_1,\ldots,e_k\}$. The conditional probability of choosing edge
$e_{k+1}$, simplifying the notation by writing $u$ for $e_{k+1}$ and
using $z^\alpha(e)$ to denote $z^\alpha_{k}$ for edge $e=(q_k,r_k)$,
is then given by
\begin{eqnarray}
P(p,u\,|\,p) & = & \frac{Z_{p,u}}{Z_{p}}\label{eq:probmulti}\\
 & = & \frac{\sum_{\alpha}q_{\alpha}\Pf(K_{p,u}^{\alpha})\prod_{e\in
 p,u}x^{\alpha}(e)}{\sum_{\alpha}q_{\alpha}\Pf(K_{p}^{\alpha})\prod_{e
\in p}x^{\alpha}(e)}\nonumber\\
 & = &
 \frac{\sum_{\alpha}q_{\alpha}\Pf([(K^{\alpha})^{-1}]_{p,u})\Pf(K^{\alpha})\prod_{e\in p,u}x^{\alpha}(e)}{\sum_{\alpha}q_{\alpha}\
Pf([(K^{\alpha})^{-1}]_{p})\Pf(K^{\alpha})\prod_{e\in p}x^{\alpha}(e)}\nonumber\\
 & = & \frac{\sum_{\alpha}q_{\alpha}\Pf(K^{\alpha})\prod_{e\in
 p,u}z^{\alpha}(e)x^{\alpha}(e)}{\sum_{\alpha}q_{\alpha}\Pf(K^{\alpha})\prod
_{e\in p}z^{\alpha}(e)x^{\alpha}(e)}\ \nonumber\\
 & = & \frac{\sum_{\alpha}\zeta_{\alpha}(p)z^{\alpha}(u)x^{\alpha}(u)}{\sum_{\alpha}\zeta_{\alpha}(p)}\,,\nonumber
\end{eqnarray}
where
\begin{equation}
\zeta_\alpha(p)=\Pf(K^\alpha)\prod_{e\in p} \frac{x^\alpha(e)z^\alpha(e)}{|x^\alpha(e)|}\ .
\end{equation}
This extra weighting quantity, $\zeta_\alpha(p)$, is not needed for
planar samples, due to cancellations, but is required here to allow
for the different $p$-dependent weightings resulting from the distinct
boundary conditions. It incorporates the weight of the whole
$K^\alpha$ matrix, the modification of those weights by the factors of
$z^\alpha(e)$ resulting from the choice of edges in $p$, and the sign
of the weights (the magnitudes are identical in each $\alpha$ for a
given choice of $p$ and hence cancel out). This weighting factor can be
updated at each stage $k$
along with the set of $V^\alpha_k$, $M^\alpha_k$, and
$A^\alpha_k$ for each $\alpha$. In the case of the periodic
lattice, these four sets of matrices are updated and used to compute the
values of $z^\alpha(e)$ to find the conditional probabilities.

\subsection{Sampling spins}

\begin{figure}
  \includegraphics[%
    width=\singfigwidth, keepaspectratio]{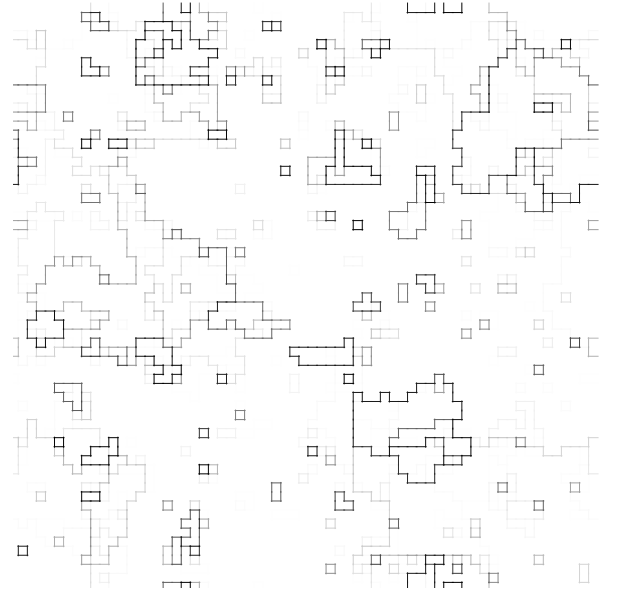}
  \caption{\label{fig:periodicSample} Relative domain walls found in
    an individual 2D Ising spin glass sample with $64^2$ spins,
    periodic boundary conditions, and unit variance Gaussian disorder,
    for temperature $T=0.16$. The bond satisfaction probabilities were
    estimated by averaging over 660 samples. As in
    Fig.\ \ref{fig:firstFish}, the lines indicate the probability of
    relative domain walls between two configurations: the darkest
    lines indicate where the bond dual to that domain wall has a
    nearly 50\% chance of opposite or equal relative orientations;
    where there is no line or a light line separating two spins, the
    two spins have a very high probability of a single relative
    orientation, either aligned or opposite.}
\end{figure}

The dimer assignments are carried out on $G$ for the periodic case
using Eq.\ (\ref{eq:probmulti}).  To finally carry out the sampling on
the torus, one first arbitrarily sets the value of an initial spin,
the spin at the upper left corner, i.e., at location $(0,0)$. The spin
at the left side of the second row, at location $(1,0)$, is fixed by
the first element of the matching for the first separator. This is the
exceptional case for this lattice where one has three choices for the
matching edge on $G$ ($(0,1)$, $(0,2)$, and $(0,6L-2)$). After this
choice has been made, the rest of the spins in the first two rows are
then assigned as in the fixed boundary case. An example of the
relative domain wall density for a $64^2$ periodic sample is displayed
in Fig.\ \ref{fig:periodicSample}. This plot shows the variance
$\mu_{ij}(1-\mu_{ij})$ in the bond satisfaction,
where $\mu_{ij}$ is the probability of a given bond being
satisfied, i.e., $s_i s_j J_{ij} > 0$.

\subsection{Running time}

We find that the number of bits required for the periodic case
increases only by a small amount, about 1\%, over the planar case for
samples of the same size. Carrying out the initial Pfaffian elimination for
single $\alpha$ for the entire sample is slower than for the planar case, as there are about
four times as many operations, but this
computation requires only a small fraction of the time in any case. However,
as the periodic case requires the maintenance of four $V_k$, $A_k$, and $M_k$ matrices,
sampling in the periodic case is slower than for the fixed boundary case.
We find that sample generation is about $5.5$ times slower for
periodic samples, compared with planar samples, for $L=16$ through $L=64$.

\section{Concluding comments}

In this paper, we have described an algorithm that generates spin
configurations for the 2D Ising spin glass, where the samples
generated are directly selected according to the equilibrium
probability distribution. This method follows from Wilson's dimer
sampling algorithm, though we have modified the matrix algebra for
speed and simplicity, and have adopted the dimer matching to the study of
the Ising spin glass. We have also generalized the method to periodic
samples.

We note that as the inverse Kasteleyn matrix contains the dimer-dimer
correlation functions along the separator, one need not carry out all
of the sampling steps to compute domain wall densities.  One can
directly examine the inverse on the separator to find the domain wall
densities on a single separator, by stopping at step \ref{step4} of
the outline in Sec.\ \ref{Sec:Wilson}.  The separator can then be
changed to compute the bond satisfaction probabilities in each row of the
sample.  Sampling configurations provides more information, but if the
bond satisfaction variance is all that is needed, this approach is
more precise and is not unreasonably slow.

This algorithm can also be used to directly and uniformly sample ground states
in the 2D $\pm J$ spin glass model. At low enough temperatures (on the order of
$T\approx 0.1$), the ground states occur frequently, as can be confirmed by
their energies being lowest or by comparison with a ground state energy found
by combinatorial optimization. The statistics of the ground state
configurations can therefore be directly sampled (by rejecting other states
when they occur), exactly, using this algorithm.

Our implementation of the sampling algorithm is efficient enough to
allow for rapid enough sampling to study finite temperature patchwork
dynamics out to patch sizes $\ell$ of at least $\ell=32$. Large numbers of samples
can be comfortably generated for $L=64$ and $T<0.01$ or for $L=128$
and $T=0.04$.  This should allow for more conclusive studies on the
Gaussian and $\pm J$ spin glass problems in two dimensions.

We thank Jan Vondrak for sharing his code for computing the partition function
of the $\pm J$ spin glass model and Simon Catterall for sharing his code for
computing Pfaffians. This work was supported in part by the NSF under
grant No.~0606424.

\end{document}